\documentclass[nofootinbib,floats,floatfix,eqsecnum,prd,aps]{revtex4}
\usepackage[latin1]{inputenc}
\usepackage[T1]{fontenc}
\usepackage[dvips]{graphicx}
\usepackage{hyperref}
\usepackage{dcolumn,epsfig,minitoc}
\usepackage{amssymb,amsmath}
\usepackage[usenames]{color}
\def\o1{$O_{1}$}
\def\o2{$O_{2}$}

\newcommand{\beq}{\begin{equation}}
\newcommand{\eeq}{\end{equation}}
\newcommand{\bea}{\begin{eqnarray}}
\newcommand{\eea}{\end{eqnarray}}

\begin{document}

\title{INFRARED BEHAVIOR OF SCALAR CONDENSATES IN EFFECTIVE HOLOGRAPHIC
THEORIES}
\author{Mariano Cadoni$^{1}$, Paolo Pani$^{2,3}$ and Matteo Serra$^{1}$}
\affiliation{$^{1}$ Dipartimento di Fisica, Universit\`a di Cagliari and INFN, Sezione di
Cagliari - Cittadella Universitaria, 09042 Monserrato, Italy.\\
$^{2}$CENTRA, Departamento de Física, Instituto Superior T\'ecnico,
Universidade T\'ecnica de Lisboa - UTL, Avenida~Rovisco Pais 1, 1049 Lisboa,
Portugal.\\
$^{3}$ Institute for Theory $\&$ Computation, Harvard-Smithsonian CfA, 60
Garden Street, Cambridge, MA, USA.}
\date{\today }

\begin{abstract}
We investigate the infrared behavior of the spectrum of  scalar-dressed, 
asymptotically Anti de Sitter (AdS) black
brane (BB) solutions of effective holographic models. These solutions
describe scalar condensates in the dual field theories. We show that for
zero charge density the ground state of these  BBs must be
degenerate with the AdS vacuum, must satisfy conformal boundary conditions
for the scalar field and it is isolated from the continuous part of the
spectrum. When a finite charge density is
switched on, the ground state is not anymore isolated and the degeneracy is
removed. Depending on the coupling functions, the new ground state may possibly 
be energetically
preferred with respect to the extremal Reissner-Nordstrom AdS BB. We 
derive several properties of BBs near extremality and at finite temperature.
As a check and illustration of our results we derive and discuss several
analytic and numerical,  BB solutions of
Einstein-scalar-Maxwell AdS gravity with different coupling functions 
and different potentials. We also discuss how our results can be used for
understanding holographic quantum critical points, in particular their
stability and the associated quantum phase transitions leading to
superconductivity or hyperscaling violation.
\end{abstract}

\author{}
\maketitle
\tableofcontents



\section{Introduction}

Holographic models have been widely used as a powerful tool to describe the
strongly coupled regime of quantum field theory (QFT) \cite{Hartnoll:2008vx,Hartnoll:2008kx,
Horowitz:2008bn,Herzog:2009xv,Hartnoll:2009sz,Charmousis:2009xr,
Cadoni:2009xm,Goldstein:2009cv,Gubser:2009qt,Goldstein:2010aw,
Charmousis:2010zz,Bertoldi:2010ca, Gouteraux:2011ce,Iizuka:2011hg,
Cadoni:2011kv,Hartnoll:2011fn,Lee:2011zzf,Kim:2012nb, Dong:2012se,
Wu:2012fk,Hashimoto:2012ti,Ammon:2012je,Iizuka:2012wt,Dey:2012rs,
Park:2012cu,Myung:2012cb,Adam:2012mw,Gouteraux:2012yr}. In the usual
holographic setup, the $4$-dimensional gravitational bulk is described by an
asymptotically anti-de Sitter (AdS) black brane (BB), generically sourced by
a (neutral or charged) scalar field and by an electromagnetic (EM) field.
The usual rules of the AdS/CFT correspondence are then used to describe a
dual $3$-dimensional QFT at finite charge (or finite chemical potential) and
where some scalar operator may acquire a nonvanishing expectation value. In
particular, when the BB solution is sourced by a nontrivial scalar field
configuration in the bulk, the dual QFT shows a (neutral or charged) scalar
condensation. 

These effective holographic theories (EHTs) have a wide range
of applications. They have been used to give a holographic description of
many interesting quantum phase transitions, such as those leading to
critical superconductivity and hyperscaling violation \cite{Hartnoll:2008vx,Hartnoll:2008kx,Charmousis:2009xr,Gubser:2009qt,
Cadoni:2009xm,Goldstein:2009cv,Dong:2012se,Cadoni:2011kv,Cadoni:2012uf,Cadoni:2012ea}%
. The properties of the QFT at small temperatures are essentially determined
by the quantum phase transition at zero temperature. Thus, understanding the
quantum critical point provides a full characterization of the
thermodynamical phase transition at small temperatures.

If one assumes, as we do in this paper, that the asymptotic geometry is the
AdS spacetime,  the dual QFT shows a universal conformal fixed point in
the ultraviolet (UV). The nontrivial dynamics therefore occurs in the
infrared (IR) region, at the corresponding critical points. In a Wilsonian
approach, EHTs should be first classified in terms of flows, driven by
relevant operators, between critical points corresponding to scale-invariant
(more generally scale-covariant) QFTs. Two other relevant characterizations
of the critical points are: $a)$ the distinction between \textsl{%
fractionalized} phases (sourced by non-zero electric flux in the IR) and 
\textsl{cohesive} phases (sourced by zero electric flux in the IR); $b)$
phases with broken and unbroken $U(1)$ symmetry \cite%
{Hartnoll:2011fn,Gouteraux:2012yr}.

Progress in the classification and understanding of IR critical points have
been achieved following various directions. In particular, it has been shown
that in the case of hyperscaling preserving and hyperscaling violating
solutions, quantum critical theories may appear as fixed lines rather than
fixed points \cite{Gouteraux:2012yr}. Hyperscaling preserving solutions
appear indeed as fixed points and correspond to AdS$_{4}$, AdS$_{2}\times
R^{2}$ and Lifshitz bulk geometries. However, hyperscaling violating
solutions are characterized by an explicit scale and therefore appear rather
as critical lines generated by changing that scale or equivalently the
charge density \cite%
{Cadoni:2012uf,Cadoni:2012ea,Gouteraux:2012yr,Dong:2012se}.

A crucial point for understanding these quantum critical points is the
presence of a scalar condensate. Indeed nontrivial configurations of
(generically charged or uncharged) scalar fields play several crucial roles:
(i) nontrivial scalar fields are dual to relevant operators that drive the
renormalization group (RG) flow from the UV fixed point to the IR critical point (or line); (ii)
scalar fields are the sources that support the IR hyperscaling violating
geometry allowing for both fractionalized and cohesive phases \cite%
{Hartnoll:2011fn,Gouteraux:2012yr}; (iii) charged scalar condensates break
the $U(1)$ symmetry and generate a superconducting phase \cite%
{Hartnoll:2011fn,Gouteraux:2012yr}.

Despite the recognized relevance of scalar condensates for describing
holographic critical points, we are far from having a complete
understanding of the physics behind them, in particular we have very few
information about their stability. For instance, one would like to
understand why at zero (and small) temperature the hyperscaling violating
phase is energetically preferred with respect to the hyperscaling preserving
phase. In this paper we will move a step forward in this direction by asking
ourselves a simple, but relevant question: what is the energy of the ground
state of a neutral asymptotically AdS BB sourced by a generic nontrivial
scalar field? We show that for zero charge density the BB ground state must
be degenerate with the AdS vacuum. This degeneracy is the result of an exact
cancellation between a positive gravitational contribution to the energy and
a negative contribution due to the scalar condensate.

Moreover, we also show that for the BB ground state the symmetries of 
the field equations force
conformal boundary conditions for the scalar field, i.e.
boundary conditions preserving the asymptotic symmetry group of the AdS
spacetime. The conformal boundary conditions correspond to dual multitrace
scalar operators driving the dynamics from the UV conformal fixed point to
the IR critical point. In the case of an IR hyperscaling violating geometry
sourced by a pure scalar field with a potential behaving exponentially, a
scale is generated in the IR. On the
other hand we will show that, in the case of pure Einstein-scalar gravity at
finite temperature $T$, the boundary conditions for the scalar are
determined by the dynamics and are, therefore, generically nonconformal.
This means that the ground state for scalar BBs is \textsl{isolated}, i.e.
it cannot be obtained as the $T\to0$ limit of finite-$T$ BBs with conformal
boundary conditions for the scalar field.

When a finite charge density $\rho$ is switched on, the degeneracy of the
ground state is removed. Because an additional degree of freedom (the EM potential) is
present, the boundary conditions for the scalar field are not anymore
determined by the dynamics. The freedom in choosing the boundary conditions
arbitrarily can be used to impose conformal boundary conditions also for BBs at finite temperature.
The ground state for scalar BBs is therefore not anymore
isolated from the continuous part of the spectrum. The coupling between the
bulk scalar and EM field determines if it is energetically preferred with
respect to the extremal Reissner-Nordstrom (RN) AdS BB.

We also derive several properties of the scalar BB near extremality
and at finite temperature. For instance, we show that scalar-dressed,
neutral (charged), BB solutions of radius $r_{h}$ (and charge density $\rho$%
) only exist for a temperature $T$ bigger than the temperature of the
Schwarzschild-AdS (Reissner-Nordstrom AdS) BB with the same $r_{h}$ (and with the same $\rho$%
).

As a check and illustration of our results we give and discuss --~both
analytically and numerically~-- several (un)charged, scalar-dressed BB
solutions of Einstein-scalar-Maxwell AdS (ESM-AdS) gravity with minimal,
nonminimal and covariant coupling functions and different potentials
(quadratic, quartic, exponential).

Finally, we also discuss the relevance of our results for understanding
holographic quantum critical points, in particular their stability and the
associated quantum phase transitions.

The structure of this paper is the following. In Sect. \ref{sect:f1} we
present the general form of the EHTs we consider. In Sect. \ref{sect:extremal}
we investigate the spectrum of this class of theories in the IR region. In
Sect. \ref{sect:neutral} we derive extremal, near-extremal and
finite-temperature BB solutions of pure Einstein-scalar gravity theories in
the case of a quadratic, quartic and exponential potential. We also derive
their thermodynamical behavior and their critical exponents. In Sect. \ref%
{sect:charged} we derive and discuss charged solutions with the scalar
minimally, nonminimally and covariantly coupled to the EM field. Finally in
Sect. \ref{sect:concluding} we end the paper with some concluding remarks
about the relevance that our results have for understanding the dual QFT,
holographic quantum critical points, in particular the stability of the latter and the
associated quantum phase transitions leading to superconductivity or
hyperscaling violation. In Appendix \ref{sect:appendix} we discuss
perturbative solutions in the small scalar field limit.

\section{Effective holographic theories}

\label{sect:f1} We consider Einstein gravity coupled to a real scalar field
and to an EM field in four dimensions: 
\begin{equation}
I=\int d^{4}x\sqrt{-g}\left[ R-\frac{1}{2}\left( \partial \phi \right) ^{2}-%
\frac{Z(\phi )}{4}F^{2}-V(\phi )-Y(\phi )A^{2}\right] .  \label{action}
\end{equation}%
where $F_{\mu \nu }=\partial _{\mu }A_{\nu }-\partial _{\nu }A_{\mu }$ is
the Maxwell field-strength. The model is parametrized by the gauge coupling
function $Z(\phi )$, by the self-interaction potential $V(\phi )$ for the
scalar field and by the coupling function $Y(\phi )$ giving mass to the
Maxwell field.

The action (\ref{action}) defines ESM theories of gravity, which are also called EHTs and are
relevant for holographic applications. EHTs have
been widely used to give a holographical description of strongly-coupled QFTs with
rich phenomenology such as quantum phase transitions, superconductivity and
hyperscaling violation \cite%
{Hartnoll:2008vx,Hartnoll:2008kx,Charmousis:2009xr,Gubser:2009qt,
Cadoni:2009xm,Goldstein:2009cv,Dong:2012se,Cadoni:2011kv,Cadoni:2012uf,
Cadoni:2012ea,Gouteraux:2012yr}. Moreover, models like (\ref{action})
generically appear, after dimensional reduction, as low-energy effective
string theories. The action (\ref{action}) can be also interpreted as an EHT
for a complex scalar field that enjoys a $U(1)$ symmetry \cite%
{Gouteraux:2012yr}. In this context the real scalar $\phi $ describes the
modulus of the charged scalar and the phase with broken (unbroken) $U(1)$
symmetry is obtained by $Y\neq 0$ ($Y=0$).

Although our considerations can be easily extended to the case $Y\neq 0$, we
will focus for simplicity on the case of unbroken $U(1)$ symmetry, $Y=0$. We
will briefly comment on the case $Y\neq0$ in Section \ref{sect:sbreak}.

We are interested in electrically charged BB solutions of the theory, i.e.
static solutions with radial symmetry for which the topology of the
transverse space is planar. Using the following parametrization for the
metric: 
\begin{equation}
ds^{2}=-\lambda (r)dt^{2}+\frac{dr^{2}}{\lambda (r)}+H^{2}(r)(dx^{2}+dy^{2}),
\label{metric}
\end{equation}%
the Einstein and scalar equations read: 
\begin{eqnarray}
\frac{H^{\prime \prime }}{H} &=&-\frac{(\phi ^{\prime })^{2}}{4},\quad
(\lambda H^{2})^{\prime \prime }=-2H^{2}V,  \label{eqsEM1} \\
(\lambda HH^{\prime })^{\prime } &=&-H^{2}\left[ \frac{V}{2}+\frac{%
ZA_{0}^{\prime }{}^{2}}{4}\right] \,,  \label{eqsEM2} \\
(\lambda H^{2}\phi ^{\prime })^{\prime } &=&H^{2}\left( \frac{dV}{d\phi }-%
\frac{A_{0}^{\prime }{}^{2}}{2}\frac{dZ}{d\phi }\right).  \label{eqsEM3}
\end{eqnarray}%
The ansatz~\eqref{metric} is very convenient, as in these coordinates
Maxwell's equations can be directly solved for $A_{0}^{\prime }$: 
\begin{equation}
A_{0}^{\prime }=\frac{\rho }{ZH^{2}},  \label{eqsEM4}
\end{equation}%
where $\rho $ is the charge density of the solution. Note that only Eqs. (%
\ref{eqsEM2}) and (\ref{eqsEM3}) depend on the EM field and only through $%
A_{0}^{\prime }$. Therefore, substituting the solution above into the
remaining field equations, we can completely eliminate the EM field and
solve Eqs.~\eqref{eqsEM1}--\eqref{eqsEM3} for $\lambda $, $H$ and $\phi $.

We will consider models for which the potential $V(\phi )$ has a maximum at $%
\phi =0$  and $Z^{\prime }(\phi =0)=0$, with the local mass of the scalar $%
m_{s}^{2}=V^{\prime \prime }(0)$ satisfying the condition $%
m_{BF}^{2}<m_{s}^{2}\leq -2/L^{2}$ and with $V(0)=-6/L^{2}$, where $m_{BF}^{2}=-9/(4L^{2})$ is the
Breitenlohner-Freedman (BF) bound~\cite{Breitenlohner:1982bm} in four
dimensions and $L$ is the AdS length\footnote{%
The results of our paper can be easily extended to the scalar-mass range $%
m_{BF}^{2}<m_{s}^{2}<m_{BF}^{2}+1/L^{2}$, where the dual CFT to is known to
be unitary.}. The presence of an extremum of $V(\phi )$ and $Z(\phi )$ at $%
\phi =0$ implies the existence of a Reissner-Nordstrom-AdS (RN-AdS) BB solution, 
\begin{equation}
\lambda =\frac{r^{2}}{L^{2}}-\frac{M}{2r}+\frac{\rho ^{2}}{4r^{2}},\quad
H=r,\quad \phi =0,  \label{sads}
\end{equation}%
which is characterized by a trivial scalar field configuration.

On the other hand well-known ``no-hair'' theorems \cite%
{Torii:2001pg,Hertog:2006rr,Cadoni:2011nq,Cadoni:2011yj} make the existence
of BB solutions endowed with a nontrivial scalar field a rather involved
question. Scalar-dressed solutions are particularly important for
holographic applications because they describe dual QFTs with a scalar
condensate.

The AdS, $r=\infty $, asymptotic behavior requires the following leading
behavior of the metric and the scalar field: 
\begin{eqnarray}
ds^{2} &=&-\frac{r^{2}}{L^{2}}dt^{2}+\frac{L^{2}}{r^{2}}%
dr^{2}+r^{2}(dx^{2}+dy^{2})  \notag  \label{bc} \\
\phi &=&\frac{O_{1}}{r^{\Delta _{1}}}+\frac{O_{2}}{r^{\Delta _{2}}},
\end{eqnarray}%
with $\Delta _{1,2}=\frac{3\mp \sqrt{9+4m_s^2L^{2}}}{2}$. Because the 
AdS spacetime is not globally hyperbolic,  this
asymptotic behavior must be supported by boundary conditions on $O_{1}$, $%
O_{2}$. Dirichlet boundary conditions that preserve the asymptotic
isometries of the AdS spacetime are $O_{1}=0$. However, in the range of
scalar masses $m_{BF}^{2}<m_{s}^{2}<m_{BF}^{2}+1/L^{2}$, boundary conditions
of the form ($f$ is a constant): 
\begin{equation}
O_{1}=fO_{2}^{\Delta _{1}/\Delta _{2}},  \label{cbc}
\end{equation}%
which preserve the conformal, asymptotic symmetries of the AdS background,
are also allowed~\cite{Hertog:2004dr}. More in general, boundary conditions
of the form 
\begin{equation}
O_{1}=W(O_{2}),  \label{bc2}
\end{equation}%
can be used. For a generic form of the function $W$ the asymptotic AdS
isometries are broken, yet an asymptotic time-like Killing vector exists and
the gravitational theory admits a dual description in terms of multitrace
deformations of CFTs \cite%
{Witten:2001ua,Aharony:2001pa,Berkooz:2002ug,Marolf:2006nd,Hartman:2006dy,Vecchi:2010dd,Hertog:2004ns}.

Apart from their UV AdS behavior, the scalar-dressed solutions of EHTs are
also characterized by their, small $r$, IR behavior. This IR behavior
is of crucial relevance for holographic applications, in particular in the
context of the AdS/Condensed Matter correspondence~\cite{Hartnoll:2009sz}.
Generically, we expect the IR regime not to be universal, but rather
determined by the infrared behavior of the potential $V(\phi)$ and of the
gauge coupling functions $Z(\phi)$, $Y(\phi)$. Nevertheless, we will
discover in the next sections some features of the IR spectrum of EHTs,
which are model-independent and related to the scaling symmetries of the UV
AdS vacuum.

Although we will be concerned with general features of EHTs, for the sake of
definiteness we will mainly focus on three classes of models with different
IR behavior of the potential $V(\phi)$:

\begin{itemize}
\item[a)] The potential has a quadratic form 
\begin{equation}
V(\phi )=-\frac{6}{L^{2}}+\frac{m_s^2\phi ^{2}}{2}.  \label{quadratic}
\end{equation}%
This corresponds to the simplest choice for the potential, which has been
widely used in holographic models. The IR regime is dominated by the
quadratic term and at $T=0$ the scalar field diverges logarithmically in the 
$r=0$, near-horizon region.

\item[b)] The potential behaves exponentially for small values of the radial
coordinate $r$. Assuming that $r=0$ corresponds to $\phi \rightarrow \infty $%
, we have in this case 
\begin{equation}
V(\phi )\sim e^{b\phi },  \label{moda}
\end{equation}%
where $b$ is a positive constant. As we shall discuss later in Sect. \ref%
{sect:exponential}, this case produces a scale-covariant solution in the IR,
corresponding to hyperscaling violation in the dual QFT.

\item[c)] The origin $r=0$ corresponds to an other extrema (a minimum) at $%
\phi=\phi_{1}$ of the potential $V(\phi)$. In this case the theory flows to
a second AdS$_{4}$ vacuum in the infrared.
\end{itemize}

The IR regime of the EHT (\ref{action}) is also characterized by the IR
behavior of the gauge coupling function $Z$. In particular, $Z$ is crucial
for determining the contribution of bulk degrees of freedom inside or
outside the event horizon to the boundary charge density. This distinction
is captured by the behavior of the electric flux in the IR 
\begin{equation}
\Phi =\left( \int_{R^{2}}Z(\phi )\tilde{F}\right)_{\mathrm{IR}},
\label{flux}
\end{equation}%
where $\tilde{F}$ is the dual Maxwell tensor. Using a terminology borrowed
from condensed matter physics, the phase with $\Phi =0$ has been called 
\textsl{cohesive} and describes dual confined gauge invariant operators. The
phase $\Phi \neq 0$ has been named \textsl{fractionalized} and describes a
dual deconfined phase \cite{Hartnoll:2011fn,Gouteraux:2012yr}. In this paper
we will consider two choices for the gauge coupling function $Z(\phi 
)$: $(1)$
a minimal coupling, $Z(\phi )=1$; $(2)$ a coupling which behaves
exponentially in the IR, $Z\sim e^{a\phi }$.

Since in the following we shall make often use of the thermodynamical
properties of the BB solutions, we find it convenient to summarize them
here. The temperature $T$, entropy $S$ and free energy $F$ of the solutions (%
\ref{metric}) are given by 
\begin{equation}
T=\frac{\lambda ^{\prime }(r_{h})}{4\pi }\,,\qquad S=4\pi \mathcal{V}
H^{2}(r_{h}),\text{ \ \ \ \ \ }F=M-TS,  \label{d1}
\end{equation}%
where $M$ is the \textit{total} mass of the solution, $\mathcal{V}$ is the
volume of the $2D$ sections of the spacetime and $r_{h}$ is the location of
the outer event horizon. 

\section{ Spectrum of Einstein-Scalar-Maxwell AdS gravity in the Infrared
region}

\label{sect:extremal} In this section we investigate general features of the
mass spectrum of ESM-AdS gravity in the IR region. Assuming the existence of
scalar-dressed BBs with AdS asymptotic behavior, the two basic questions in
this context are about the existence and features of the $T=0$ extremal
state and of the states near-extremality. We will treat separately the EM
charged and uncharged cases. We will first consider the theory with zero
charge density ($Z=Y=0$ in the action (\ref{action})), i.e. a vanishing
Maxwell field (Einstein-scalar AdS gravity). Later, we will extend our
considerations to the case of finite charge density.

\subsection{ Einstein-Scalar AdS gravity}

\label{sect:esads} A nontrivial point is the determination of the total mass 
$M$ (i.e. the energy) of the BB solution. As discussed in Ref. \cite%
{Hertog:2004ns}, the usual definition of energy in AdS diverges whenever $%
O_{1}\neq 0$ (with a divergent term proportional to $r$). This is because
the backreaction of the scalar field causes certain metric components to
fall off slower than usual. However, this divergent term is exactly canceled
out if one considers that for $O_{1}\neq 0$ there is an additional scalar
contribution to the surface terms which determines the mass. 

Using the
Euclidean action formalism in the case $m_{s}^{2}=-2/L^{2}$ the total mass
turns out to be \cite{Hertog:2004ns} 
\begin{equation}
M=M_{G}+\frac{\mathcal{V}}{L^{4}}\left[O_{1}O_{2}+P(O_{1})\right] ,
\label{mass}
\end{equation}%
where $M_{G}$ is the gravitational contribution to the mass, we have chosen
the following boundary conditions for the scalar: $O_{2}=O_{2}(O_{1})$, and $%
P(O_{1})=\int O_{2}(O_{1})dO_{1}$.

In the following we will need an expression for the mass when $m_{s}$ is in
the range of values considered in this paper, $-9/4<m_s^2 L^2\leq -2$.
Furthermore, working with the parametrization of the metric given by Eq. (%
\ref{metric}), it is useful to express the total mass $M$ in terms of the
coefficient of the $1/r$ term in the $r=\infty $ expansion of the metric
functions. To derive such an expression we use the Euclidean action
formalism of Martinez et al. \cite{Martinez:2004nb}. Using the
parametrization of the metric (\ref{metric}) the gravitational and scalar
part of the variation of the boundary terms are given respectively by \cite%
{Martinez:2004nb}: 
\begin{eqnarray}
\delta I_{G} &=&\frac{2\mathcal{V}}{T}\left[ (HH^{\prime }\delta \lambda
-\lambda ^{\prime }H\delta H)+2\lambda H(\delta H^{\prime })\right]
|_{r_{h}}^{\infty },  \notag  \label{s3} \\
\delta I_{\phi } &=&\frac{\mathcal{V}}{T}H^{2}\lambda \phi ^{\prime }\delta
\phi |_{r_{h}}^{\infty }.
\end{eqnarray}%
From the definition of the free energy $F=M-TS$, taking into account that $%
I_{\phi }|_{r_{h}}=0,\,S=I_{G}|_{r_{h}}$ and from $F=-IT$, it follows 
\begin{equation}
M=TS-TI=-T(I_{G}^{\infty }+I_{\phi }^{\infty }).  \label{mass2}
\end{equation}

To calculate the mass $M$ (\ref{mass2}) we need the subleading terms in the $%
r=\infty $ expansion of the metric (\ref{bc}). By means of a translation of
the radial coordinate $r$, the asymptotic expansion of the solution can be
put in the general form: 
\begin{eqnarray}  \label{expansion_inf}
\lambda &=&\frac{r^{2}}{L^{2}}+pr^{\beta }-\frac{m_{0}}{2r}+{\mathcal{O}}({%
r^{\beta -1}}),  \notag  \label{gae} \\
H^{2} &=&\frac{r^{2}}{L^{2}}+qr^{\alpha }+\frac{s}{r}+{\mathcal{O}}({%
r^{\alpha -1}}),  \notag \\
\phi &=&\frac{O_{1}}{r^{\Delta _{1}}}+\frac{O_{2}}{r^{\Delta _{2}}}+{%
\mathcal{O}}({r^{-\Delta _{1}-1}}),
\end{eqnarray}%
where $p,q,\alpha ,\beta ,s$ are constants. Inserting this expansion in the
field equations one gets (at the first and second subleading order) the
following relations between the constants: 
\begin{equation}
\beta =\alpha =2(1-\Delta _{1}),\quad p=q=\frac{\Delta _{1}O_{1}^{2}}{%
4L^{2}(1-2\Delta _{1})},\quad s=-\frac{\Delta _{1}\Delta _{2}O_{1}O_{2}}{%
6L^{2}}.  \label{k11}
\end{equation}%
Substituting Eq. (\ref{gae}) into (\ref{s3}) and using $p=q$, we obtain 
\begin{eqnarray}
\delta I_{G}^{\infty } &=&-\frac{\mathcal{V}}{TL^{2}}\left( \delta
m_{0}+6\delta s-2\delta p(\beta -1)r^{\beta +1}\right) \,,  \label{s4} \\
\quad \delta I_{\phi }^{\infty } &=&-\frac{\mathcal{V}}{TL^{4}}\left( \Delta
_{1}O_{1}\delta O_{1}r^{\beta +1}+\Delta _{1}O_{1}\delta O_{2}+\Delta
_{2}O_{2}\delta O_{1}\right) \,.
\end{eqnarray}%
Notice that both the gravitational and the scalar contribution to the mass
contain a term which diverges as $r^{\beta +1}$. Using Eq.~(\ref{k11}) one
easily finds that the two divergent terms cancel out in $\delta I^{\infty
}=\delta I_{G}^{\infty }+\delta I_{\phi }^{\infty }$ . Finally, we obtain
the total mass of the solution 
\begin{equation}
M=-TI^{\infty }=\frac{\mathcal{V}}{L^{2}}\left( m_{0}+\frac{(\Delta
_{1}-\Delta _{2})}{L^{2}}\int dO_{2}W(O_{2})+\frac{\Delta _{2}(1-\Delta _{1})%
}{L^{2}}O_{2}O_{1}\right) ,  \label{s7}
\end{equation}%
where we have parametrized the boundary conditions for the scalar in terms
of the function $O_{1}=W(O_{2})$. It is also useful to split the total mass
into the gravitational and scalar contributions $M_{G}$ and $M_{\phi }$,
arising separately from the two terms in Eq.~(\ref{s7}):

\begin{equation}
M_{G}=\frac{\mathcal{V}}{L^{2}}\left( m_{0}-\frac{\Delta _{1}\Delta _{2}}{%
L^{2}}O_{1}O_{2}\right) ,\quad M_{\phi 
}=\frac{\mathcal{V}}{L^{4}}\left[
\Delta _{1}O_{1}O_{2}+(\Delta _{2}-\Delta _{1})P(O_{1})\right] ,  \label{c3}
\end{equation}%
where $P(O_{1})$ is defined as in Eq. (\ref{mass}). One can easily check
that the previous equations reproduce correctly Eq. (\ref{mass}) in the case 
$m_{s}^{2}=-2/L^{2}$, i.e. $\Delta _{1}=1,\Delta _{2}=2$.

Let us now investigate general features of the mass spectrum of ES-AdS
gravity in the IR region. In particular, assuming the existence of
scalar-dressed BBs with AdS asymptotic behavior, we wish to characterize the
features of the $T=0$ extremal state and of the near-extremal states.

In the uncharged case a general, albeit implicit, form of the solution for
the metric function $\lambda $ in a generic ES-AdS gravity theory has been
derived in Ref.~\cite{Cadoni:2011nq}: 
\begin{equation}
\lambda =H^{2}-C_{1}H^{2}\int \frac{dr}{H^{4}},  \label{d3}
\end{equation}%
where $C_{1}$ is an integration constant. The equation above implies that if
an extremal $T=0$ hairy BB solution exists, this must have $C_{1}=0$, i.e. $%
\lambda =H^{2}$. We can prove this statement by the following argument.
Differentiating the equation above and using Eqs.~\eqref{d1}, we find the
following relation between the temperature and the entropy density $\mathcal{%
S}$ of the solution: 
\begin{equation}
T=\frac{\lambda ^{\prime }(r_{h})}{4\pi }=\frac{\left( 2\lambda HH^{\prime
}-C_{1}\right) _{r_{h}}}{{\mathcal{S}}}.
\end{equation}%
%
%
%
%
%
%
Therefore we get $C_{1}=\left[ 2\lambda HH^{\prime }\right] _{r_{h}}-%
\mathcal{S}T$. An extremal solution satisfies $T=0$ and $\lambda (r_{h})=0$.
Assuming that $H$ and $H^{\prime }$ are finite at the horizon (to avoid
curvature singularities), the existence of an extremal solution imposes $%
C_{1}=0$, i.e. 
\begin{equation}
\lambda =H^{2}.  \label{r8}
\end{equation}%
Note that this argument applies both when the entropy of the extremal
solution is vanishing or when it is finite.

Obviously, an extremal uncharged solution with AdS asymptotics (besides the
trivial AdS vacuum) may not exist. For the moment, we assume such a solution
exists and derive a general and very important result. We shall prove that
if such a solution exists it must have zero energy, i.e. \textit{must be
degenerate with the AdS vacuum}.

To prove this statement, let us first show that every scalar-dressed
solution with AdS asymptotics, which is characterized by $\lambda =H^{2}$,
requires necessarily conformal boundary conditions (\ref{cbc}) for the
scalar field. The field equations~\eqref{eqsEM1}--\eqref{eqsEM4} with $%
\rho=0 $ and the metric (\ref{metric}) are invariant under the scale
transformation $r\rightarrow \mu r,\,\lambda \rightarrow \mu ^{2}\lambda
,\,t\rightarrow \mu ^{-2}t$. In the extremal case, the asymptotic expansion (%
\ref{gae}) implies that full solution ($\lambda =H^{2}$ and $\phi $) is
invariant under this scale transformation if $O_{1,2}$ scale as follows: $%
O_{1}\rightarrow \mu ^{\Delta _{1}}O_{1},\,O_{2}\rightarrow \mu ^{\Delta
_{2}}O_{2}$, which in turn implies the conformal boundary condition~%
\eqref{cbc}.

We can now calculate the mass (\ref{s7}) of the extremal solution, which has 
$\lambda =H^{2}$, hence $m_{0}=-2s$. We get, 
\begin{equation}
M=\frac{\mathcal{V}}{L^{4}}\left[ (\Delta _{2}-\Delta _{1})P(O_{1})+\Delta
_{1}(1-\frac{2}{3}\Delta _{2})O_{2}O_{1}\right] ,  \label{s9}
\end{equation}%
where $P(O_{1})$ is defined as in Eq. (\ref{mass}). From Eq. (\ref{s9}) it
follows almost immediately that for conformal boundary conditions (\ref{cbc}%
) the mass $M$ vanishes.

This is an important and extremely nontrivial result. It means that in
ES-AdS gravity with no EM field, if an extremal scalar-dressed BB solution
exists then the AdS$_{4}$ vacuum of the theory must necessarily be
degenerate. Physically, this degeneration is a consequence of the fact that
the scalar condensate gives a negative contribution to the energy. Therefore
we may have configurations in which the positive gravitational energy is
exactly canceled by the negative energy of the scalar condensate. This
cancellation is a consequence of the conformal symmetry of the extremal
solution; it necessarily occurs because the extremality condition $\lambda
=H^{2}$ forces the conformal boundary conditions (\ref{cbc}).

It is also important to notice that the argument leading to the degeneracy
of the $T=0$ ground state holds true also when a condition much weaker than
Eq. (\ref{r8}) is satisfied: 
\begin{equation}
\lambda =H^{2}+{\mathcal{O}}(r^{-2}).  \label{r9}
\end{equation}%
In fact the mass (\ref{s7}) and the scaling arguments leading to the
conformal boundary conditions for the scalar field depend only on terms up
to ${\mathcal{O}}(r^{-1})$ and are completely insensitive to higher order
terms in $1/r$.

Let us now consider near-extremal solutions. We assume that the theory
allows for scalar-dressed BBs at finite temperature with AdS asymptotics. In
the next section, we shall prove the existence of finite temperature
solutions, by constructing AdS-BBs, numerically, for three classes of ES-AdS
gravity models. 

The BB spectrum near-extremality can be investigated by
considering the $T\rightarrow 0$ limit of the finite $T$ solutions. However,
one can show that this $T\rightarrow 0$ limit is singular. In order to prove
the statement we expand the fields in the near-horizon region, 
\begin{equation}
\lambda =\sum_{n=1}^{\infty }a_{n}(r-r_{h})^{n},\quad H=\sum_{n=0}^{\infty
}b_{n}(r-r_{h})^{n},\quad \phi =\sum_{n=0}^{\infty }c_{n}(r-r_{h})^{n}\,.
\label{h1}
\end{equation}%
At first order we get for $b_{0}\neq 0$ 
\begin{equation}
b_{2}=-\frac{b_{0}}{4}c_{1}^{2},\quad b_{1}a_{1}=-\frac{b_{0}}{2}%
V(c_{0}),\quad a_{1}c_{1}=\left(\frac{dV}{d\phi }\right)_{c_{0}},
\label{l01}
\end{equation}

whereas the temperature of the dressed solutions, from Eq.~\eqref{d1},
becomes 
\begin{equation}
T=\frac{a_{1}}{4\pi }=-\frac{b_{0}V(c_{0})}{8\pi b_{1}}\,.
\label{Tuncharged}
\end{equation}

Because in the case under consideration ($V$ has a maximum) the potential $V$
is limited from above ($V(\phi )\leq -6/L^{2}$)$,$ the $T\rightarrow 0$
limit can only reached by letting $b_{0}\rightarrow 0$. But on the other
hand from the third equation in (\ref{l01}) it follows immediately that $%
a_{1}=0$ is a singular point of the perturbative expansion (\ref{h1}) unless 
$(dV/d\phi )_{c_{0}}=0$ (corresponding to the Schwarzschild-anti de Sitter
(SAdS) BB). Thus the $T\rightarrow 0$ limit is a singular point of the
perturbation theory. It should be stressed that this result has been derived
by first considering the near-horizon limit, then taking $T\rightarrow 0$.
In section \ref{sect:pert} we will see what happens if the two limits are
taken in the reversed order.

Note that the above results are strictly true only if one considers AdS
solutions with negative mass squared for the scalar field. If the scalar
potential has a local minimum at $\phi =0$, then our argument above does not
apply. This is for instance the case with the class of models studied in
Ref.~\cite{Cadoni:2011yj} which, however, turn out not to have BB solutions
with AdS asymptotics.

The singularity of the $T\to0$ limit in the near-horizon perturbation
theory, indicates that the ground state (\ref{r8}) is isolated, i.e. it
cannot be reached as the $T\to0$ limit of finite$-T$ scalar BB solutions.
This conclusion can be also inferred by reasoning on the $r=\infty$ boundary
conditions for the scalar field. We have previously shown that the
symmetries of the field equations together with Eq. (\ref{r8}) force
conformal boundary conditions (\ref{cbc}) for the scalar field. On the other
hand, one can easily show that in the case of $T$ finite, the field
equations together with the conditions for the existence on an event horizon
imply boundary conditions of the form (\ref{bc2}), hence in general
nonconformal boundary conditions. In fact, the field equations~\eqref{eqsEM1}%
--\eqref{eqsEM4} have the following symmetries: 
\begin{eqnarray}  \label{rescaling}
&&r\rightarrow kr,\quad t\rightarrow kt,\quad L\rightarrow kL,\quad
H\rightarrow kH  \notag  \label{fed2} \\
&&r\rightarrow kr,\quad \lambda \rightarrow k^{2}\lambda ,\quad t\rightarrow
k^{-1}t,\quad A_{0}\rightarrow kA_{0}  \label{fed2b} \\
&&H\rightarrow kH,\quad (x,y)\rightarrow k^{-1}(x,y),  \notag \\
&&\lambda \rightarrow k\lambda ,\quad t\rightarrow k^{-1}t,\quad
H^{2}\rightarrow k^{-1}H^{2},\quad L\rightarrow kL\,,\quad
A_{0}^{2}\rightarrow k^{-1}A_{0}^{2}  \notag
\end{eqnarray}
These symmetries can be used to fix all but one parameter in the
perturbative expansion (\ref{h1}). The solutions become in this way a
one-parameter family of solutions. The near-horizon expansion (\ref{h1})
depends on a single free parameter, which can be chosen to be $r_{h}$. For
each value of $r_{h}$, we can extract the two functions $O_{1}(r_{h})$ and $%
O_{2}(r_{h})$, which define implicitly the boundary condition $%
O_{1}=W(O_{2})$.

It follows that in general the finite$-T$ solution require boundary
conditions for the scalar, which are different from the conformal ones
required for the ground state (\ref{r8}). Therefore, the solution (\ref{r8})
is generically isolated, i.e. it cannot be reached as the $T\to0$ limit of
finite$-T$ scalar BB solutions.

It should be stressed that the above feature is a key general feature of the
BB solutions of AdS Einstein-scalar gravity which holds true also for the
numerical solutions discussed in the next sections. If one assumes an
analytic expansion close to the horizon, an asymptotically AdS behavior at
infinity and if one requires the existence of hairy BB solutions, then the
boundary conditions at infinity cannot be arbitrarily imposed but are
determined by the field equations. These boundary conditions will have the
form (\ref{bc2}), with the function $W$ determined by the dynamics. In the
dual QFT the function $W$ characterizes the scalar condensate. Thus, the
particular form of the condensate is determined by the gravitational
dynamics. It is obvious that this is true only in the case of pure
Einstein-scalar gravity. For instance it does not hold for electrically
charged solutions\footnote{%
It does not hold also for black hole solutions of ES-AdS gravity, i.e. for
solutions which spherical horizons. This is because in this case the field
equations are not anymore invariant under the full set of transformations (%
\ref{rescaling}).}. In this latter case the near-horizon solution has always
more than one free parameter, that can be fixed by prescribing some boundary
condition for the scalar field.

We can also compare the temperature of the dressed solution of radius $r_{h}$
with the temperature $T_{0}$ of the SAdS BB with  the same radius. We can use
Eqs. (\ref{rescaling}) to set $r_{h}=L$,$\,b_{1}=c_{1}=L^{-1},b_{0}=1$, so
that the only free parameter is $c_{0}=\phi (r_{h})$ and the temperature
becomes $-8\pi T=LV(c_{0})$. We have therefore 
\begin{equation}
T-T_{0}=8\pi L(V(0)-V(c_{0})\,).  \label{k5}
\end{equation}%
%
%
%
%
%
%
In the case under consideration, $V(\phi )$ has a local maximum at $\phi =0$%
, so that $V(0)\geq V(c_{0})$. Therefore, we obtain $T>T_{0}$ for \textit{any%
} finite temperature solution. That is, there exists a critical temperature
given by the temperature of the SAdS BB: $T_{0}=\frac{3r_{h}}{4\pi L^{2}}$
such that scalar-dressed solutions of the same radius $r_{h}$ only exist
when $T>T_{0}$.


\subsection{Einstein-Scalar-Maxwell AdS gravity}

Let us now consider the EM charged case, i.e. a finite charge density in the
dual QFT. In general, one expects that the finite charge density will remove
the degeneracy of the $T=0$ extremal state we have found in the uncharged
case. This can be shown explicitly. Indeed, when $\rho \neq 0$, the field
equations imply 
\begin{equation}
\frac{\rho ^{2}}{ZH^{2}}+2\lambda {H^{\prime }}^{2}+2\lambda HH^{\prime
\prime }=H^{2}\lambda ^{\prime \prime }\,,
\end{equation}
which is solved by Eq.~\eqref{d3} only when the charge is vanishing. In
particular, $\lambda =H^{2}$ is not a solution of the equation above when $%
\rho \neq 0$. Moreover in the charged case Eq. (\ref{d3}) becomes \cite%
{Cadoni:2011nq} 
\begin{equation}
\lambda =H^{2}\left[1-C_{1}\int \frac{dr}{H^{4}}+\rho ^{2}\int dr\left( 
\frac{1}{H^{4}}\int \frac{dr}{Z H^{2}}\right)\right] .  \label{d9}
\end{equation}

By using the same procedure leading to Eq. (\ref{r8}), we get that the
extremal solution in the EM charged case is attained for 
\begin{equation}
C_{1}=\rho ^{2}\left( \int \frac{dr}{ZH^{2}}\right) _{r_{h}}.  \label{p1b}
\end{equation}%
Because $C_{1}\neq 0$, not even the weaker condition (\ref{r9}) is satisfied
in the charged case. This implies that Eq. (\ref{s9}) does not hold if $\rho
\neq 0$. In general, the mass of the extremal scalar-dressed solution will
be different from the mass of the extremal RN-AdS solution, so that the
degeneration of the $T=0$ ground state in the EM charged case is removed.

Notice that in the charged case the $T=0$ solution is not necessarily forced
to have conformal boundary conditions for the scalar field. In fact, the
argument used for the uncharged case is based both on
the relation $\lambda =H^{2}$ and on the scale symmetries of the field
equations. Both do not hold anymore at finite charge density. Nevertheless,
in this case the presence of an additional field (the EM potential $A_{\mu }$%
) allows to choose arbitrary boundary conditions for the scalar. As
discussed in the previous section, the boundary conditions are not anymore
imposed by the dynamics of the system as in the uncharged case. It follows
that in the charged case the $T=0$ ground state is not anymore isolated but
can be reached continuously as the $T=0$ limit of finite temperature scalar
dressed BB solutions.

For what concerns the BB spectrum near extremality, the results we have
found in the uncharged case still hold in the charged case. In fact the
first and the third equation in (\ref{l01}) are not modified by the
nonvanishing charge, whereas the second becomes $a_{1}b_{1}=-(Z^{-1}(c_{0})%
\rho ^{2}+2b_{0}^{4}V(c_{0}))/4b_{0}^{3}$. We obtain the temperature: 
\begin{equation}
T=\frac{a_{1}}{4\pi }=-\frac{Z^{-1}(c_{0})\rho ^{2}+2b_{0}^{4}V(c_{0})}{%
16\pi b_{1}b_{0}^{3}}\,.  \label{Tcharged}
\end{equation}%
Also here, we can compare the temperature of a scalar-dressed BB with that
of the RN BB with the same charge $\rho $ and radius $r_{h}$. One easily
finds that Eq. (\ref{k5}) still holds for the charged case and that $%
T>T_{0}^{RN}$ for \textit{any} finite temperature solution, where the
critical temperature $T_{0}^{RN}$ is given by 
\begin{equation}
T_{0}^{RN}=\frac{12r_{h}^{4}-L^{2}Z^{-1}(c_{0})\rho ^{2}}{16\pi
r_{h}^{3}L^{2}}\,.  \label{Tc}
\end{equation}%
Scalar dressed EM charged solutions of the same radius $r_{h}$ and charge of
the RN-AdS solution exist only for $T>T_{0}^{RN}$.

An important issue when dealing with finite EM charge density is the
characterization of the phase as fractionalized or cohesive~\cite%
{Hartnoll:2011fn,Gouteraux:2012yr}. For the generic theory (\ref{action})
with $Y\neq 0$, this characterization will depend on the IR behavior of both 
$Z(\phi)$ and $Y(\phi)$. However, one can easily show that in the case of
unbroken $U(1)$ symmetry, $Y=0$, only the fractionalized phase may exist. In
fact using Eq. (\ref{eqsEM4}) into Eq. (\ref{flux}) one easily finds $%
\Phi\sim \rho$.

To summarize, the following interesting picture emerges for the IR spectrum
of scalar-dressed BB solutions of ESM-AdS gravity with $%
m_{BF}^{2}<m_{s}^{2}<m_{BF}^{2}+1/L^{2}$. If a scalar-dressed, 
neutral,  extremal
solution exists at $\rho =0$, it must necessarily be degenerate with the AdS
vacuum. This is due to a precise cancellation of the contributions to the
total energy from the gravitational and scalar part and, in turn, it is due
to the conformal symmetries of the boundary theory. Moreover, the $%
T\rightarrow 0$ limit of finite-$T$ BB solution is singular and the ground
state is isolated from the continuous part of the spectrum.

When an EM charge is switched on, the degeneracy of the ground state is
removed and the ground state can be reached continuously as the $T\to0$
limit of finite$-T$ solutions. Scalar dressed uncharged (EM charged)
solutions of the same radius $r_{h}$ (and charge) of S-AdS (RN-AdS) solution
exist only for $T>T_{0}$ ($T>T_{0}^{RN}$). Cohesive phases may exist only
when the $U(1)$ symmetry is broken. In the $U(1)$ symmetry-preserving case
only the fractionalized phases are allowed.

Our results are fairly general and only assume the existence of scalar
dressed solutions, which has to be investigated numerically. In the next two
sections we will show that the picture above is realized for three wide
classes of models with quadratic, quartic and exponential potentials $V(\phi
)$ and for two classes of gauge couplings ($Z=1$ and $Z\sim e^{a\phi}$).
Numerical computations confirms the degeneracy of the ground state in the
uncharged case and the peculiarity of the $T\rightarrow 0$ limit of
finite-temperature scalar-dressed BB solutions. We will discuss separately
the EM neutral and charged solutions.

\section{Neutral solutions}

\label{sect:neutral}

\subsection{Quadratic potential}

\label{sect:quadratic} In this section we will construct numerical solutions
of ES-AdS gravity models with the quadratic potential (\ref{quadratic}) and
we shall check the validity of the general results of Section \ref%
{sect:extremal}. The case of a quadratic potential is the simplest possible
choice and it is therefore our first example. Moreover, this is the usual
choice for models describing holographic superconductors. We will come back
to this point later in Sect. \ref{sect:sbreak}.

\subsubsection{Extremal solutions}

The near-horizon behavior of the extremal solution of the model (\ref%
{quadratic}) with an EM field covariantly coupled to a charged scalar field
has been derived in Ref.~\cite{Horowitz:2009ij}. The near-horizon, extremal
solution of a pure Einstein-scalar gravity model (both the EM and the charge
of the scalar field are zero) can be obtained as a particular case of the
solution given in Ref.~\cite{Horowitz:2009ij}. In the gravitational gauge
used in Ref.~\cite{Horowitz:2009ij}, we have 
\begin{equation}
ds^{2}=-g(\hat{r})e^{-\chi }dt^{2}+\frac{d\hat{r}^{2}}{g(\hat{r})}+\hat{r}%
^{2}(dx^{2}+dy^{2}),  \label{metric1}
\end{equation}%
and with our normalization for the kinetic term of the scalar field, the
solution reads 
\begin{equation}
ds^{2}=\frac{d\hat{r}^{2}}{g_{0}\hat{r}^{2}(-\ln \hat{r})}+\hat{r}%
^{2}(-dt^{2}+dx^{2}+dy^{2}),\quad \phi =2\sqrt{2}(-\ln \hat{r})^{1/2},\quad
g_{0}=-\frac{2m_{s}^{2}}{3}.  \label{hr}
\end{equation}%
The near-horizon, extremal solution (\ref{hr}) can be written in the gauge (%
\ref{metric}) by a suitable reparametrization of the radial coordinate. We
get 
\begin{equation}
\lambda =H^{2}=e^{-\frac{g_{0}X^{2}(r)}{2}},\quad \phi =-\sqrt{2g_{0}}%
\,X(r),\quad r=\sqrt{\frac{\pi }{g_{0}}}+\int^{X}dte^{-\frac{g_{0}t^{2}}{4}},
\label{hr1}
\end{equation}%
where the last equation defines implicitly the function $X(r)$. We note that
also in these coordinates the horizon is located at $r={0}$.

\begin{figure*}[htb]
\begin{center}
\begin{tabular}{cc}
\epsfig{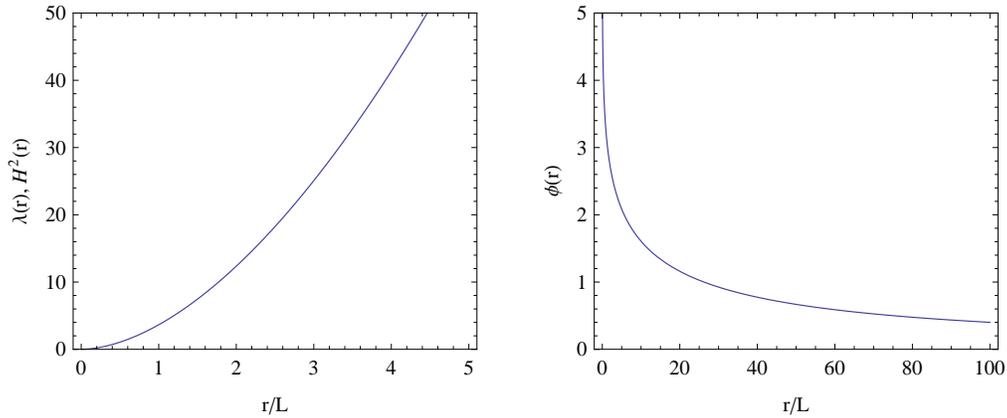} & 
\end{tabular}%
\end{center}
\caption{The metric functions $\protect\lambda =H^{2}$ (left) and the scalar
field $\protect\phi $ (right) as functions of $r/L$, in the extremal case,
for a quadratic potential with $m^2_{s}L^2=-2$.}
\label{quadr_profiles}
\end{figure*}

The global, extremal, solution interpolating between the near-horizon
behavior (\ref{hr1}) and the asymptotic AdS behavior (\ref{gae}) has to be
found numerically. We have integrated the field equations numerically for
several values of $m_{s}^{2}$. In all cases we have found $\lambda =H^{2}$,
which implies the conformal boundary condition (\ref{cbc}). Indeed the total
mass $M$ of the scalar-dressed solution is zero. In Fig.~\ref{quadr_profiles}
we show the profiles of the metric functions and the scalar field for $%
m_{s}^{2}=-2/L^{2}$.

\subsubsection{Finite-temperature solutions}

\label{section:finitetemp} Let us now consider BB solutions of
Einstein-scalar theory at finite temperature. Again we have to construct
global solutions, which interpolate between the asymptotic AdS expansion
given by Eq. (\ref{gae}) and a near-horizon expansion as in Eq.~\eqref{h1}.

We have constructed these solutions numerically, starting from the
near-horizon solution above and integrating outwards to infinity, where the
asymptotic behavior of the solution is AdS$_{4}$. In Fig.~\ref%
{fig:quadratic_profiles} we show an example of the metric and scalar
profiles and of the function $O_{2}(O_{1})$ in the case $m_{s}^{2}=-2/L^{2}$%
\footnote{In Fig. \ref%
{fig:quadratic_profiles} and in all the figures we  show in this 
paper  all the dimensional quantities ($O_{1,2}, F,{\cal{F}}, c,T$) are 
normalized with appropriate powers of the AdS length $L$} . In the large $O_{1}$ limit, our data are well fitted by $O_{2}\sim
-0.57O_{1}^{2}$, which is consistent with the conformal boundary condition~%
\eqref{cbc}. However, for small values of $O_{1}$ the behavior reads $%
O_{2}\sim -0.36O_{1}$ and the global behavior interpolates between these two
asymptotic regimes. Therefore, the function $O_{2}(O_{1})$ does not
generically satisfy the conformal boundary conditions (\ref{cbc}). This is a
general statement that we have verified also for different choices of the
parameters and different models. This fact confirms that extremal solution
are isolated from finite temperature solutions.

As expected, solutions dressed with scalar hairs only exist \textit{above} a
certain critical temperature $T_{0}=3/(4\pi )$ and a critical mass $M_{0}$
which correspond to the temperature and mass of the Schwarzschild-AdS BB
after a rescaling that sets $r_{h}=L=1$. This is shown in Fig.~\ref%
{fig:quadratic_mass}, where we present the \textit{total} mass $M$ of the
solutions as a function of the temperature $T$. The absence of dressed
solutions (irrespectively of the boundary conditions $O_{2}(O_{1})$) for $%
T<T_{0}$ confirms numerically the existence of the critical temperature $%
T_{0}$.

\begin{figure*}[htb]
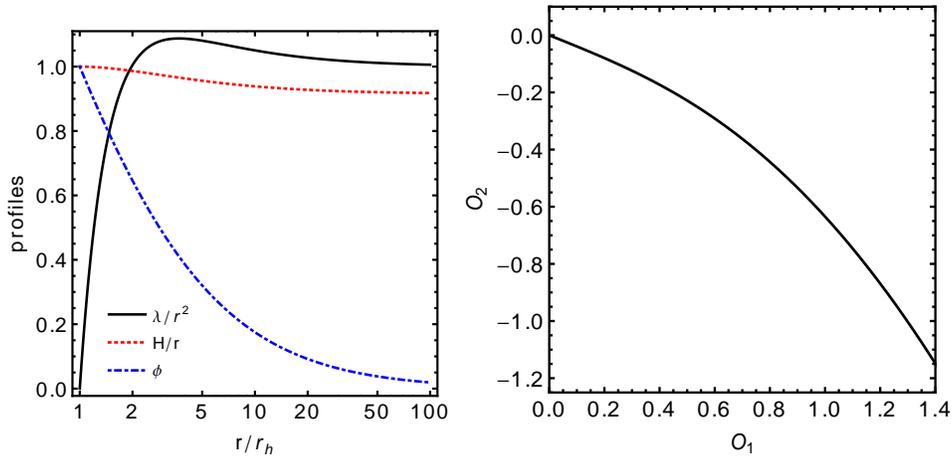

\begin{center}
\begin{tabular}{cc}
\epsfig{file=figure/quadratic_profiles.eps,height=6cm,angle=0,clip=true} & %
\epsfig{file=figure/O2O1.eps,height=6cm,angle=0,clip=true}%
\end{tabular}%
\end{center}
\caption{Left panel: metric and scalar profiles as functions of the
(nonrescaled) coordinate $r$ for a quadratic potential with $m^2_{s}L^2=-2$%
. Right panel: the function $O_2=O_2(O_1)$ for the same model.}
\label{fig:quadratic_profiles}
\end{figure*}

\begin{figure*}[tbh]
\begin{center}
\begin{tabular}{cc}
\epsfig{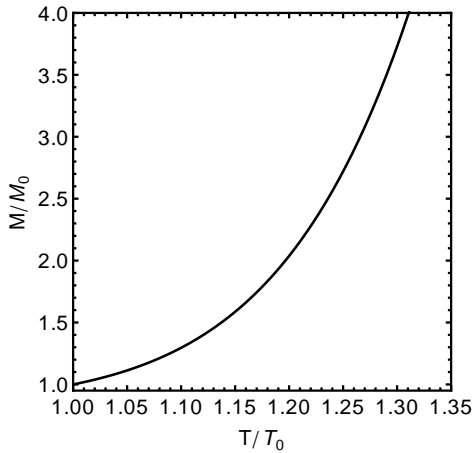} & 
\end{tabular}%
\end{center}
\caption{Total mass $M$ as a function of the temperature $T$ for a quadratic
potential with $m^2_{s}L^2=-2$.}
\label{fig:quadratic_mass}
\end{figure*}

\subsection{ Quartic potentials}

\label{sect:quartic} In this section we will check numerically the results
of Section~\ref{sect:extremal}, and the validity of the picture that has
emerged from our results, in the case of a theory with a potential $V(\phi )$
having the behavior described as type $c)$ in Section~\ref{sect:f1}, i.e. a
theory with a IR fixed point. As an example of such a theory we take the
quartic potential, 
\begin{equation}
V(\phi )=\Lambda ^{4}\phi ^{4}-\frac{\hat{m}^{2}}{2}\phi ^{2}-\frac{6}{L^{2}}%
.  \label{quartic}
\end{equation}%
This potential has the typical mexican hat form with a maximum at $\phi =0$
with $V(0)=-6/L^{2},\,V^{\prime \prime }(0)=-\hat{m}^{2}$ and a minimum at $%
\phi _{12}=\pm \frac{\hat{m}}{2\Lambda ^{2}}$ with $V(\phi _{1})=-6/l^{2}=-%
\hat{m}^{4}/(16\Lambda ^{4})-6/L^{2},\,V^{\prime \prime }(\phi _{1})=2\hat{m}%
^{2}$. The potential is invariant under the discrete transformation $\phi
\rightarrow -\phi $, so that we will just consider $\phi \geq 0$. The theory
allows for two AdS$_{4}$ vacua: an UV AdS at $\phi =0$ (corresponding to $%
r=\infty $), with AdS length $L$ and with squared mass of the scalar given
by $-\hat{m}^{2}$, and an IR $AdS_{4}$ at $\phi =\phi _{1}$ (corresponding
to $r=0$) with AdS length $l$ and with squared mass of the scalar given by $2%
\hat{m}^{2}$. Again, we focus on $-9/4<-\hat{m}^{2}L^{2}\leq -2$.

\subsubsection{Extremal solutions}

A scalar-dressed, extremal solution of the kind discussed in the previous
section would represent a flow between an UV AdS$_{4}$ and an IR AdS$_{4}$.
Let us first investigate numerically the existence of such a solution. If it
exists we know from the results of the previous section that it must have
zero mass, i.e. it must be degenerate with the (UV) AdS vacuum. In order to
construct such solution numerically we need its perturbative expansion in
the UV (near $r=\infty $) and in the IR (near horizon, $r=0$). The UV
expansion is given by Eq. (\ref{gae}). For what concerns the near-horizon $%
r=0$ expansion, the field equations (\ref{eqsEM1})-(\ref{eqsEM3}) give
instead 
\begin{equation}
\lambda =\frac{r^{2}}{l^{2}}-\frac{\gamma ^{2}}{12l^{4}}r^{4}+{\mathcal{O}}%
(r^{6}),\,\,H=\frac{r}{l}-\frac{\gamma ^{2}}{24l^{3}}r^{3}+{\mathcal{O}}%
(r^{5}),\,\,\phi =\phi _{1}+\frac{\gamma }{l}r+{\mathcal{O}}({r^{2}}),
\label{j1}
\end{equation}%
where $\gamma $ is an arbitrary constant. Moreover, Eq. (\ref{eqsEM3})
constrains the possible values of the parameter $\hat{m}$ in Eq. (\ref%
{quartic}), $\hat{m}^{2}=2/l^{2}$. Introducing a dimensionless
parametrization for $\Lambda $ in Eq.~(\ref{quartic}), $\Lambda ^{-4}=kl^{2}$%
, one finds that the restriction on $\hat{m}^{2}$ implies 
\begin{equation}
0<k<\frac{8}{3},\quad \frac{l^{2}}{L^{2}}=1-\frac{k}{24}.  \label{f1}
\end{equation}

We have integrated the field equations numerically, starting from $r\sim 0$
outwards to infinity. When $\phi \geq 0$, regular solutions only exist for $%
\gamma <0$. These solutions interpolate between the $r=\infty $ AdS behavior
(\ref{gae}) and the near horizon solution (\ref{j1}).

\begin{figure*}[tbh]
\begin{center}
\begin{tabular}{cc}
\epsfig{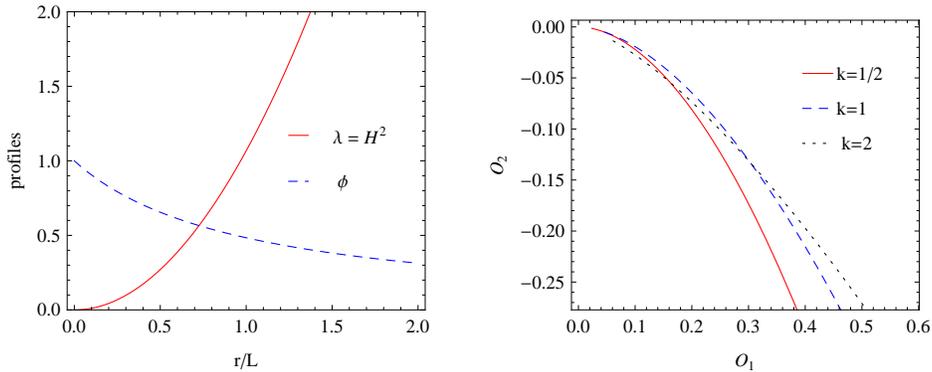} & 
\end{tabular}%
\end{center}
\caption{Left panel: profiles of the metric functions and the scalar field,
in the extremal case, for a quartic potential with $k=1$ and $\protect\gamma %
=-1$. Right panel: the function $O_{2}(O_{1})$ for three different values of 
$k$.}
\label{fig:quartic_extreme}
\end{figure*}

In Fig.~\ref{fig:quartic_extreme} we show the profiles of the metric
functions and of the scalar for $k=1$, and the function $O_{2}(O_{1})$
(obtained by varying the free parameter $\gamma $) for selected values of $k$%
. Again we have found that $\lambda =H^{2}$, which implies the conformal
boundary condition (\ref{cbc}) and that the total mass $M$ of the
scalar-dressed solution is vanishing.

\bigskip

\bigskip

\subsubsection{Finite-temperature solutions}

Using the same method described in Sect. \ref{section:finitetemp} we have
constructed, numerically, dressed BB solutions at finite temperature for
models with the potential (\ref{quartic}). We have generated global BB
solutions for $m_{s}^{2}L^{2}=-2$ and for several values of the parameter $%
\Lambda $. These solutions interpolate between the near-horizon expansion (%
\ref{h1}) and the asymptotic AdS$_{4}$ form. In Fig.~\ref%
{fig:quartic_profiles} we show an example of the metric and scalar profiles
and the function $O_{2}(O_{1})$ for the case $m_{s}^{2}L^{2}=-2$ and for
some selected values of $\Lambda $. As it is clear from Fig.~\ref%
{fig:quartic_profiles}, the function $O_{2}(O_{1})$ displays a universal
linear behavior at small $O_{1}$, which already confirms that the boundary
conditions are not conformal for any value of $\Lambda $. In addition, for
larger values of $O_{1}$ the slope of $O_{2}(O_{1})$ depends on the quartic
coupling.

\begin{figure*}[tbh]
\begin{center}
\begin{tabular}{cc}
\epsfig{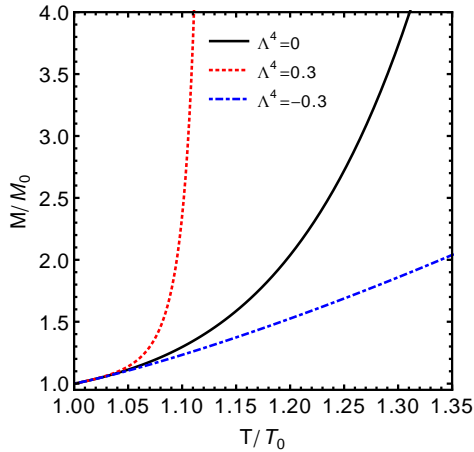} & 
\end{tabular}%
\end{center}
\caption{Total mass $M$ as a function of the temperature $T$ for a quartic
potential and for selected values of the parameter $\Lambda$.}
\label{fig:quartic_mass}
\end{figure*}

In Fig. \ref{fig:quartic_mass} we show the total mass of the solution as a
function of the temperature for fixed horizon radius $r_{h}=1$ and $L=1$. As
expected the dressed solutions exist only for $T>T_{0}$, confirming
numerically the existence of the critical temperature $T_{0}$. It should be
noticed that we have generated the numerical finite-temperature solutions
for values of the parameters $m_{s}^{2}$ and $\Lambda $, which are different
from those used to generate the extremal solutions. The reason for this
choice is a numerical instability of the solutions for positive values of $%
m_{s}^{2}$.

\begin{figure*}[htb]
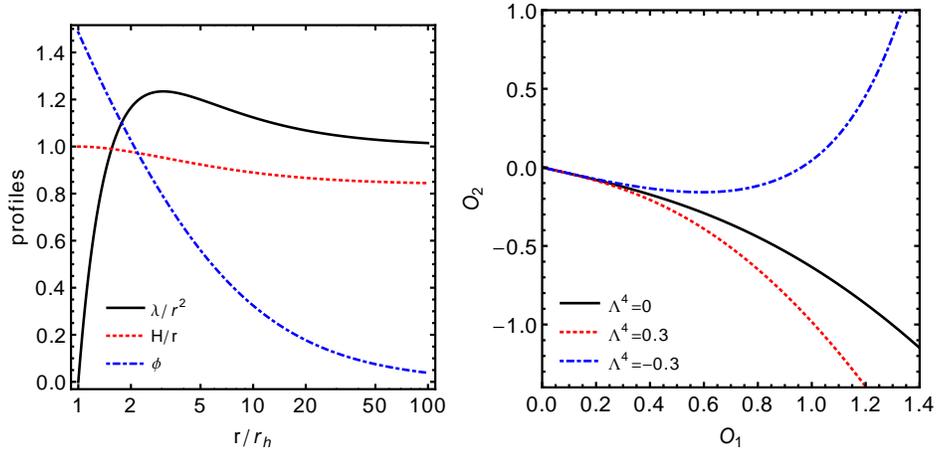

\begin{center}
\begin{tabular}{cc}
\epsfig{file=figure/quartic_profiles.eps,height=6cm,angle=0,clip=true} & %
\epsfig{file=figure/O2O1_quartic.eps,height=6cm,angle=0,clip=true}%
\end{tabular}%
\end{center}
\caption{Left panel: metric and scalar profiles as functions of the
(nonrescaled) coordinate $r$ for a quartic potential with $m^2_{s}L^2=-2 $
and $\Lambda^4=0.3$. Right panel: the functions $O_2=O_2(O_1)$ for different
values of $\Lambda$. }
\label{fig:quartic_profiles}
\end{figure*}

\subsection{Exponential potentials}

\label{sect:exponential} 
In this section we will investigate the case of a
theory with a potential $V(\phi )$ having the behavior described as type $b)$
in Section \ref{sect:f1}, i.e. the potential behaves exponentially $\sim
e^{b\phi }$ for $\phi \rightarrow \infty $ (corresponding to $r=0$). 

We search for scalar-dressed BB solutions that smoothly interpolate between
an asymptotic AdS spacetime and a near-horizon scale-covariant metric. In
the dual QFT they correspond to a flow between an UV fixed point and
hyperscaling violation in the IR. In general these interpolating solutions
cannot be found analytically but have to be computed numerically. To be more
concrete in the following we will focus on a class of models defined by the
potential 
\begin{equation}
V(\phi )=-\frac{2}{b^{2}L^{2}}[\cosh (b\phi )+3b^{2}-1].  \label{pot}
\end{equation}%
This potential is such that the mass of the scalar is independent from the
parameter $b$, $m_{s}^{2}=-2/L^{2}$. Moreover, it contains as particular
cases $b=1/\sqrt{3}$, $b=1$ models emerging from string theory
compactifications, for which analytical solutions are known \cite%
{Hertog:2004dr,Cadoni:2011nq}.

\subsubsection{Extremal solutions}

The leading near-horizon behavior of the extremal solutions can be captured
by approximating the potential in the $\phi \rightarrow \infty $ region with
the exponential form $V(\phi )=-(1/b^{2})e^{b\phi }$. In this case the field
equations (\ref{eqsEM1})-(\ref{eqsEM3}) give \cite{Cadoni:2011nq} 
\begin{eqnarray}
\lambda &=&\alpha _{0}\left( \frac{r}{r_{-}}\right) ^{w},\quad H=\left( 
\frac{r}{r_{-}}\right) ^{w/2},\quad \phi =\phi _{0}-bw\ln \left( \frac{r}{%
r_{-}}\right) ,  \label{ex} \\
\,\alpha _{0} &=&\frac{e^{b\phi _{0}}r_{-}^{2}}{b^{2}w(2w-1)},\quad w=\frac{2%
}{1+b^{2}}.  \notag
\end{eqnarray}%
Notice that $\alpha _{0}>0$ requires $w>1/2$. This restricts the parameter
range to $1/2<w<2$ ($0<b^{2}<3$). This ansatz provides an exact solution to
the equations of motion with an exponential potential $-(1/b^{2})e^{b\phi }$
but only the leading near-horizon, extremal, behavior of the solutions with $%
V(\phi )$ generic. Solution (\ref{ex}) is \textsl{scale covariant}, i.e. the
metric transforms under rescaling with a definite weight: 
\begin{equation}
r\rightarrow kr,\quad (t,x,y)\rightarrow k^{1-w}(t,x,y),\quad
ds^{2}\rightarrow k^{2-w}ds^{2}.  \label{sc}
\end{equation}

The extremal solution (\ref{ex}) contains an IR length-scale $r_{-}$.
However, in the case of neutral BB the scaling transformations (\ref{sc}) 
may change this scale. The metric part of the solution
is scale-covariant whereas the leading $\log r$ term of the scalar is left
invariant. The only  parameter that flows when IR length-scale 
$r_{-}$ is changed, is the constant mode $\phi_0$ of the scalar.

To reduce the number of independent parameters, we can exploit the
symmetries of the field equations previously discussed [cf. Eqs. (\ref%
{rescaling})] to fix $L=1$ and $\phi _{0}=0$ in Eq. (\ref{ex}). So we can
start from the more simple ansatz containing only one free parameter $r_{-}$.

Starting from this scaling behavior near the horizon and imposing an AdS
behavior (\ref{bc}) for the metric and the scalar field at infinity, we have
integrated numerically the field equations with a potential given by Eq. (%
\ref{pot}), with different values of the parameter $0<b<\sqrt{3}$. We have
found BB solutions with scalar hair, that interpolate between the
near-horizon (\ref{ex}) and asymptotic (\ref{bc}) behavior.

\begin{figure*}[htb]
\begin{center}
\begin{tabular}{c}
\epsfig{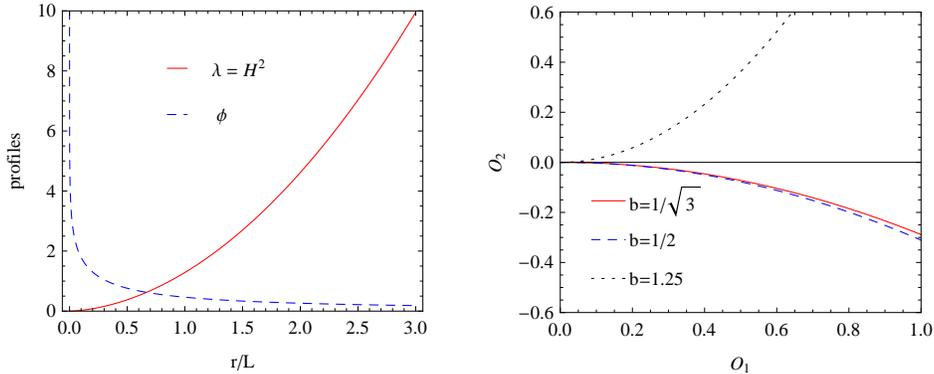}%
\end{tabular}%
\end{center}
\caption{Left panel: profiles of the metric functions and the scalar field,
in the extremal case, for the potential (\protect\ref{pot}) with $b=1/2$ and 
$r_{-}=1$. Right panel: the function $O_{2}(O_{1})$ for three different
values of $b$.}
\label{expextreme}
\end{figure*}

In Fig. \ref{expextreme} we show the metric functions and the scalar field
of these extremal BBs for $b=1/2$ and the function $P(O_{1})$ (obtained by
varying the free parameter $r_{-}$) for different values of the parameter $b$%
. Also in this case we have checked numerically that $\lambda =H^{2}$ \ and
that the conformal boundary conditions $P(O_{1})\sim O_{1}^{2}$ are
satisfied. We have also explicitly checked that the mass of the extremal
solution vanishes.

For the two cases $b=1/\sqrt{3}$ and $b=1$ the extremal solutions are known
analytically \cite{Cadoni:2011nq}. They are respectively given by 
\begin{eqnarray}
\lambda &=&H^{2}=\frac{(r+r_{-})^{\frac{1}{2}}}{L^{2}}r^{\frac{3}{2}},\quad
\phi =-\frac{\sqrt{3}}{2}\log \left( \frac{r}{r+r_{-}}\right) ,  \notag
\label{fq} \\
\lambda &=&H^{2}=\frac{r+r_{-}}{L^{2}}r,\quad \phi =-\log \left( \frac{r}{%
r+r_{-}}\right) ,
\end{eqnarray}%
where $r_{-}$ is a constant. From solutions (\ref{fq}) one can easily derive
the function $P(O_{1})$ defining the asymptotic boundary conditions for the
scalar field. We have $P=(2/\sqrt{3})O_{1}^{2}$ for $b=1/\sqrt{3}$ and $%
P=O_{1}^{2}$ for $b=1$.

In order to compare these analytical solutions with those obtained
numerically, we need to eliminate a linear term in the asymptotic behavior
of $\lambda (r)$. Taking into account this translation, we have checked
explicitly that our numerical solutions with $b=1/\sqrt{3}$ and $b=1$ and
the numerical calculated functions $P$ exactly reproduce the analytical
results. In general, the proportionality factor $f$ depends on the
value of $b$. We observe that for $b<1$ $f$ is negative, for $b=1$ $f=0$,
while for $b>1$ $f$ becomes positive, as shown in Fig. \ref{expextreme}. 

\subsubsection{Solutions at finite temperature}

\label{sect:s2}

Following the same method as the one used in the previous subsections, one
can generate generic hairy BB solutions with AdS asymptotics at finite
temperature, i.e. solutions interpolating between the near-horizon (\ref{h1}%
) and the AdS (\ref{bc}) behavior. We have generated numerically these BB
solutions and found, as in the case of quartic potential discussed above,
that for every value of the parameter $b$ in the allowed range, they exist
only for $r_{h}\geq 1$. This implies the existence of a critical temperature 
$T_{0}$ below which only the SAdS BB exists.

\begin{figure*}[htb]
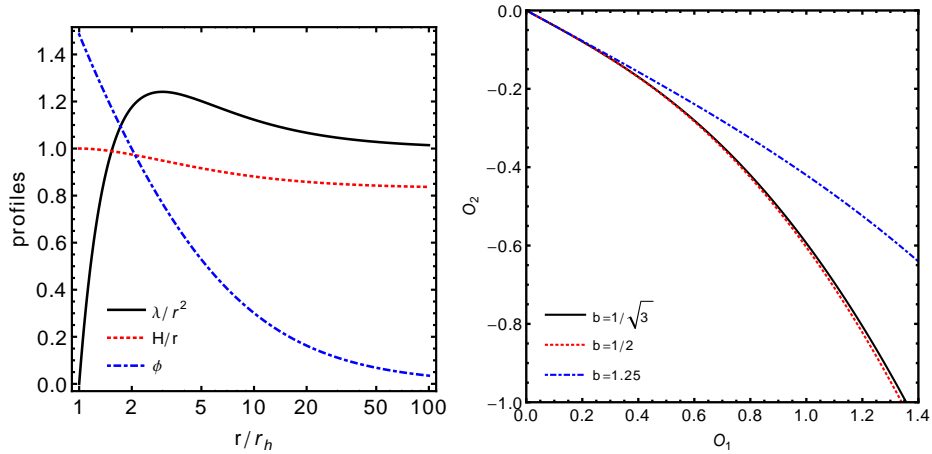

\begin{center}
\begin{tabular}{cc}
\epsfig{file=figure/cosh_profiles.eps,height=6cm,angle=0,clip=true} & %
\epsfig{file=figure/cosh_O2O1.eps,height=6cm,angle=0,clip=true}%
\end{tabular}%
\end{center}
\caption{Left panel: metric and scalar profiles as functions of the
(nonrescaled) coordinate $r$ for the potential~\eqref{pot} with $b=1/2$.
Right panel: the functions $O_2=O_2(O_1)$ for different values of $b$. }
\label{fig:cosh_profiles}
\end{figure*}

A summary of our results is presented in Fig.~\ref{fig:cosh_profiles}, which
is qualitatively similar to Fig.~\ref{fig:quartic_profiles} for the case of
quartic scalar potential. Again we have verified that the function $O_2(O_1)$
does not define conformal boundary conditions (\ref{cbc}) for the scalar,
i.e. the extremal solutions are isolated from those at finite temperature. 

\subsection{Perturbative solutions near-extremality}

\label{sect:pert} In Sect. \ref{sect:esads} we have seen that the $%
T\rightarrow 0$ limit of finite-temperature BB solution is singular and that
the ground state (\ref{r8}) is isolated from the continuous part of the
spectrum. A way to gain information about the behavior near-extremality is
to consider separately the near-horizon and near-extremal expansion. In
general the two limits do not commute. In this section we will perform this
perturbative analysis for the potential (\ref{pot}). Similar results can be
obtained for other classes of potentials.

We look for perturbative solutions of the field equations (\ref{eqsEM1})-(%
\ref{eqsEM3}) in the near-extremal, near-horizon regime. The near-extremal
regime is obtained by expanding the metric functions $\lambda $ and $H$ and $%
\phi $ in power series of an extremality parameter $m$, with $m\rightarrow 0$
when the temperature $T\rightarrow 0$ (or the BB radius $r_{h}\rightarrow 0$%
). On the other hand the near-horizon regime is obtained by expanding the
metric functions and the scalar field in power series of $r-r_{h}$. Because
in general the two limits $m\rightarrow 0$ and $r\rightarrow r_{h}$ do not
commute we have to consider separately the two cases.

\subsubsection{$r\to r_{h},\, r_{h}\to 0$}

We first expand $\lambda $ and $H$ and $\phi $ in powers of $m$: 
\begin{equation}
\lambda (r)=\sum_{n=0}^{\infty }\lambda _{n}(r)m^{n},\quad
H(r)=\sum_{n=0}^{\infty }H_{n}(r)m^{n},\quad \phi (r)=\sum_{n=0}^{\infty
}\phi _{n}(r)m^{n}.  \label{k1}
\end{equation}%
For small BB radius $r_{h}<<L$ (or equivalently small $T$, i.e. $T<<1/L$) we
can truncate in the perturbative expansion (\ref{k1}) to first order in $m$.
At leading order we find that $\lambda _{0}$, $H_{0}$ and $\phi _{0}$ must
satisfy the same field equations (\ref{eqsEM1})-(\ref{eqsEM3}). At
subleading order we find instead 
\begin{eqnarray}  \label{k2}
H_{1}^{\prime \prime } &=&-\frac{1}{4}\left( 2H_{0}\phi _{0}^{\prime }\phi
_{1}+(\phi _{0}^{\prime })^{2}H_{1}\right) ,  \label{h1q} \\
\left( 2\lambda _{0}H_{1}+H_{0}^{2}\lambda _{1}\right) ^{\prime \prime }
&=&4 \left[ \lambda _{0}\left( H_{0}H_{1}\right)^{\prime }+\lambda _{1}H_{0}H_{0}^{\prime }\right] \\
\left( 2\lambda _{0}H_{1}+H_{0}^{2}\lambda _{1}\right) ^{\prime \prime }
&=&-2\left( \phi _{1}H_{0}^{2}\frac{dV(\phi _{0})}{d\phi }+2H_{0}H_{1}V(\phi
_{0})\right) \\
\left( \lambda _{0}H_{0}^{2}\phi _{1}^{\prime }+2\lambda _{0}H_{1}\phi
_{0}^{\prime }+\lambda _{1}H_{0}^{2}\phi _{0}^{\prime }\right) ^{\prime }
&=&2H_{0}H_{1}\frac{dV(\phi _{0})}{d\phi }+H_{0}^{2}\phi _{1}\frac{%
d^{2}V(\phi _{0})}{d\phi ^{2}}  \label{h2q}
\end{eqnarray}%
A solution of Eqs.~(\ref{h1q})-(\ref{h2q}) can be obtained by setting $\phi
_{1}=H_{1}=0$, so that they reduce to 
\begin{equation}
\left( H_{0}^{2}\lambda _{1}\right) ^{\prime \prime }=0,\quad \left(
H_{0}^{\prime }H_{0}\lambda _{1}\right) ^{\prime }=0,\quad \left(
H_{0}^{2}\phi _{0}^{\prime }\lambda _{1}\right) ^{\prime }=0.  \label{k3}
\end{equation}%
Equations (\ref{eqsEM1})-(\ref{eqsEM3}) for the near-extremal leading order
functions $\lambda _{0},H_{0},\phi _{0}$ can be now solved as a near-horizon
expansion in powers of $r$. The leading term in this expansion being
obviously given by Eq. (\ref{ex}), 
\bea
\lambda _{0}(r)&=&\left( \frac{r}{r_{-}}\right) ^{w}\sum_{n=0}^{\infty }\alpha
_{n}\left( \frac{r}{r_{-}}\right) ^{n},\quad H_{0}(r)=\left( \frac{r}{r_{-}}%
\right) ^{\frac{w}{2}}\sum_{n=0}^{\infty }\beta _{n}\,\left( \frac{r}{r_{-}}%
\right) ^{n},\nonumber\\ \phi _{0}(r)&=&-bw\ln \frac{r}{r_{-}}+\sum_{n=0}^{\infty
}\gamma _{n}\,\left( \frac{r}{r_{-}}\right) ^{n}.  \label{k6}
\eea%

For each order in the $r$-expansion we can then determine the corresponding
term $\lambda _{1}^{(n)}$ for $\lambda _{1}$ by solving Eqs. (\ref{k3}). One
could worry about compatibility of the three equations (\ref{k3}). However,
one can easily realize that for $H_{0}^{2}=c_{1}r^{l}$, the system (\ref{k3}%
) is always solved by $\lambda _{1}=c_{2}r^{-l+1}$ with $c_{1,2}$ constant.
This follows from the first equation in (\ref{eqsEM1}), which implies $%
H_{0}^{\prime }\propto 1/r$. The leading order in the near-horizon expansion
involves $w,\alpha _{0},\beta _{0},\gamma _{0}$. The symmetry of the field
equations (\ref{eqsEM1})-(\ref{eqsEM3}) under a rescaling of $H$ allows to
fix $\beta _{0}=1$, whereas as expected $w,\alpha _{0}$ turn out to be given
as in Eq. (\ref{ex}). At this order Eqs. (\ref{k3}) give in turn 
\begin{equation}
\lambda _{1}^{(0)}\propto r^{-w+1}.  \label{k9}
\end{equation}%
At the $n-$th order in the near-horizon expansion we find $\lambda
_{1}^{(n)}\propto r^{-w-n+1}$. The form of the near-extremal solution is
therefore given by, 
\begin{equation}
\lambda =\lambda _{0}+\frac{m}{r^{w-1}}\left( \sum_{n=0}^{\infty }\frac{%
\epsilon _{n}}{r^{n}}\right) +\mathcal{O}(m^{2}),\quad H=H_{0},\quad \phi
=\phi _{0},  \label{n1}
\end{equation}%
where $\lambda _{0},H_{0},\phi _{0}$ are given by Eqs. (\ref{k6}). Assuming $%
m<0$ in the previous equation, we find that at leading order the relation
between $m$ and $r_{h}$ is $m\propto r_{h}^{2w-1}$. Notice that this is an
expansion in $1/r$. This means that terms with higher $n$ give smaller
contributions for $r\rightarrow \infty $.

\subsubsection{$r_{h}\to 0$ ,\,$r\to r_{h}$}

This limit has been already discussed in section \ref{sect:esads}. The
expansion in powers of $(r-r_{h})$ is given by Eq. (\ref{h1}) and at leading
order the field equations (\ref{eqsEM1})-(\ref{eqsEM3}) give the relations (%
\ref{l01}) involving the parameters $a_{1,2},b_{0,1,2},c_{0,1}$. At the next
to leading order we have three more parameters $a_{3},b_{3},c_{2}$ and three
more relations. We are therefore left with 4 independent parameters $%
b_{0},c_{0},a_{1},r_{h}$. As previously discussed, the field equations have
the symmetries~\eqref{rescaling} so that $r_{h}$ is the only independent
parameter. In principle, one can now expand $%
a_{n}(r_{h}),b_{n}(r_{h}),c_{n}(r_{h})$ in powers of $r_{h}$, substitute in
Eq. (\ref{h1}) and reorganize it as the power expansion in $m$ given by Eq. (%
\ref{k1}). By retaining only the linear terms in $m$ one could then compare
the result with Eq. (\ref{n1}). Unfortunately, this is a very cumbersome
task. Indeed, terms of order ${\mathcal{O}}(r_{h})$ are generated at any
order in the near-horizon expansion (\ref{h1}). The problem has to be solved
numerically. Numerically, one can look for global solutions interpolating
between the near-horizon behavior (\ref{h1}) with a given $r_{h}$ and the
AdS asymptotic solution (\ref{bc}). 
There is no guarantee that the solutions obtained in this way match Eq. (\ref%
{n1}). This is because the two limits $r\rightarrow r_{h}$ and $%
r_{h}\rightarrow 0$ do not commute.

\subsubsection{Near-extremal numerical solutions}

We have generated numerically, for the case of the potential (\ref{pot}),
the solutions interpolating between the AdS asymptotic behavior (\ref{bc})
and the near-extremal regime given by Eq.(\ref{n1}). In Fig. \ref%
{exp_nearextreme} (left panel) we can see the profiles of the metric
functions and the scalar field for $b=1/\sqrt{3}$ and $r_{h}=10^{-2}$
(corresponding to $m=-10^{-4}$).

\begin{figure*}[htb]
\begin{center}
\begin{tabular}{c}
\epsfig{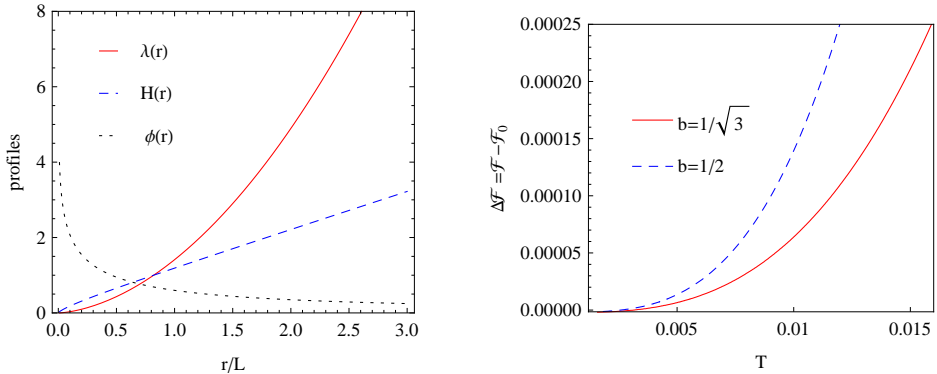}%
\end{tabular}%
\end{center}
\caption{Left panel: interpolating solutions between AdS at infinity and the
near-extreme regime given by Eq. (\protect\ref{n1}) with $b=1/\protect\sqrt{3%
}$ and $r_{h}=10^{-2}$. Right panel: difference between the free energy
density of the near-extremal BB solution and the free energy density of the $%
SAdS $ BB, for two values of $b$. }
\label{exp_nearextreme}
\end{figure*}
\bigskip

We see that although the $T\to0$ limit of the near-horizon perturbation
theory is singular and isolated, global solutions obtained interpolating the
near-horizon, near-extremal behavior (\ref{n1}) with AdS$_{4}$ exist also
for $T>0$. This is a manifestation of the noncommutativity of the
near-horizon and near extremal limit. From the point of view of perturbation
theory, the $T=0$ singularity means that the perturbative series in $m$ (\ref%
{k1}) do not converge and that solutions (\ref{n1}) are only perturbative
solutions valid for $r_{h}<<L$.

The hairy near-extremal solutions shown in Fig.~\ref{exp_nearextreme}
describe small thermal perturbations of the extremal solution, but they do
not describe the small-$T$ limit of finite-temperature solutions. These
results confirm that the ground state solutions (\ref{r8}) are not smoothly
connected to the finite-$T$ solutions, because of the existence of the
discontinuity.

Perturbative solutions in the small scalar-field limit can be also
constructed. These kind of solutions are described in the Appendix.

\subsection{Hyperscaling violation and critical exponents}

The extremal $T=0$ hairy solutions found in Sect. \ref{sect:exponential} for
the case of the potential (\ref{pot}) describe a flow between the
near-horizon regime (IR) hyperscaling violating regime and an asymptotic AdS
fixed point. From a QFT point of view, this translates into a
hyperscaling-violation phase in the IR and a scaling-preserving phase in the
UV. We can characterize the holographic features of this flow by giving the
scaling exponents in the conformal (AdS) phase and nonconformal
(hyperscaling violating) phase. The IR behavior is dictated by Eqs. (\ref{ex}%
). The UV metric is instead that of AdS$_{4}$.

To describe hyperscaling violation in four dimensions it is useful to
consider the following parametrization of the scale covariant metric:%
\begin{equation}
ds^{2}=r^{\theta -2}(-r^{-2(z-1)}dt^{2}+dx_{i}^{2}+dr^{2}),  \label{hv}
\end{equation}
where $\theta $ is the hyperscaling violation parameter and $z$ the dynamic
critical exponent. Under the following rescaling of the coordinates: 
\begin{equation*}
t\rightarrow \lambda ^{z}t,\text{ \ \ }x_{i}\rightarrow \lambda x_{i},\text{
\ \ }r\rightarrow \lambda r,\text{ }
\end{equation*}
the metric (\ref{hv}) transforms as:%
\begin{equation*}
\text{\ \ }ds\rightarrow \lambda ^{\theta /2}ds.
\end{equation*}
Moreover, this scaling transformation determines the following scaling
behavior for the free energy:%
\begin{equation}
F\sim T^{\frac{2-\theta +z}{z}}.  \label{free}
\end{equation}


By a simple redefinition of the radial coordinate and a rescaling of the
coordinates, we can write the metric (\ref{ex}) in the form (\ref{hv}). We
obtain: 
\begin{equation}
ds^{2}=r^{\frac{w}{1-w}}(-dt^{2}+dx_{i}^{2}+dr^{2}).  \label{dwh}
\end{equation}%
Comparing Eq. (\ref{dwh}) with Eq. (\ref{hv}), we can easily extract the
parameters $\theta $ and $z$ of our solution: 
\begin{equation}
z=1,\text{ \ \ }\theta =\frac{2-w}{1-w}.  \label{par}
\end{equation}%
Notice that we are now using dimensionless coordinates $r,t,x_{i}$, so that
the IR length-scale $r_{-}$ drops out from the solution. While the value $%
z=1 $ of the dynamic critical exponent is largely expected for uncharged
solutions, we see that $\theta \leq 0$ for $1<w\leq 2$ and $\theta >2$ for $%
1/2<w<1$, while $\theta $ diverges for $w=1$ (recall that in our case $%
1/2<w\leq 2$). This is in agreement with the null energy conditions for the
stress-energy tensor, which require, for $z=1$ and in the general case of $%
d+2$ dimensions, either $\theta \leq 0$ or $\theta \geq d$. Trivially, the
parameters of\ the UV AdS conformally invariant solution are $z=1,\text{\ }%
\theta =0$.

From Eq. (\ref{free}), substituting the (\ref{par}), we get that the free
energy scales as:%
\begin{equation}
F\sim T^{\frac{1-2w}{1-w}}.  \label{fre}
\end{equation}%
We see from Eq. (\ref{fre}) that the exponent of $T$ is negative for $0<w<1$
or, equivalently, when $\theta >2$. So in this case the free energy diverges
for $T\rightarrow 0$ and the corresponding phase is always unstable. 

\subsection{Thermodynamics of the near-extremal solutions}

The hairy near-extremal solutions discussed above can be interpreted as
small thermal fluctuations of extremal $T=0$ hairy BBs. The thermodynamical
features of these BB solutions -- in particular the free energy and the
specific heat -- will provide important information about the stability of
the ground state. Properties such as the scaling exponents are determined by
the behavior of the system at the quantum critical point, namely by the $T=0$
scale-covariant extremal near-horizon solution (\ref{ex}). On the other hand
the stability properties are global features and they must be investigated
using the global $T\neq 0$ solutions.

By Eqs.~\eqref{d1}, the temperature and the entropy density of the
near-horizon, near-extremal solution (\ref{n1}) are given at leading 
order by: 
\begin{equation}
T=\frac{2w-1}{4\pi }\alpha _{0}r_{h}^{w-1},\quad {\mathcal{S}}=\frac{(4\pi
)^{\frac{2w-1}{w-1}}}{[\alpha _{0}(2w-1)]^{\frac{w}{w-1}}}T^{\frac{w}{w-1}}.
\label{ther1}
\end{equation}

Notice that in these subsections we are using dimensionless coordinates, so
that the  IR length-scale $r_{-}$ drops out from our formulae, as in Eq. (%
\ref{dwh}), and we set $L=1$. Temperature and entropy density are therefore
also dimensionless.

The scaling exponent of the entropy becomes negative when $1/2<w<1$
(corresponding to $1<b^{2}<3$), implying a negative specific heat and the
corresponding solutions are therefore unstable. This is in agreement with
the results of the previous subsection concerning the scaling of the free
energy for $w<1$. Moreover, in this case small values of the temperature
correspond to high values of the horizon $r_{h}$ and of the parameter $m$,
so that we cannot obtain near-extremal solutions (in the sense of small
temperature solutions) with $r_{h}<<1$, which is the range of validity of
the perturbative solutions (\ref{n1}).

For what concerns the entropy density and free energy density $\mathcal{F}%
_{0}$ of the SAdS BB we have 
\begin{equation*}
\mathcal{S}_{0}=\frac{\left( 4\pi \right) ^{3}}{9}T^{2},\quad \mathcal{F}%
_{0}=-\left( \frac{4\pi }{3}\right) ^{3}T^{3}.
\end{equation*}

We have derived numerically the free energy of the numerical near-extremal
solutions as a function of the temperature, for $T<<1$ . In Fig. \ref%
{exp_nearextreme}{\ (right panel) we show the behavior of the free energy
density of the hairy BB solution compared with the free energy density of
the SAdS BB for two selected values of the parameter }$b$ (both such that $%
1<w<2$), and {for small values $T<<1$ of the temperature. We observe that
the scalar-dressed solutions are energetically disfavored against the SAdS
BB. This result can also be verified analytically by comparing }$\mathcal{F}$%
{$_{0}$ with the free energy density }$\mathcal{F}${\ of the hairy
near-extremal solution, which can be expressed as a function of the
temperature using Eq.~(\ref{d1}). }




\subsection{Thermodynamics of the finite temperature solutions}

We have also computed the free-energy $F$ and the specific heat $c$ of the
finite-temperature numerical scalar-dressed solutions derived in the
previous sections for the case of the quartic (\ref{quartic}) and
exponential (\ref{pot}) potential. The results are shown in Fig.~\ref%
{fig:energy} where we plot $\Delta F/F_{0}$ and $c$ as a function of the
temperature, with $\Delta F=F-F_{0}$. The free energy $F$ is always larger
than that of the corresponding Schwarzschild-AdS BB at same temperature and
the specific heat is negative. Hence, these solutions are energetically
disfavored against the undressed ones. 
\begin{figure*}[tbh]
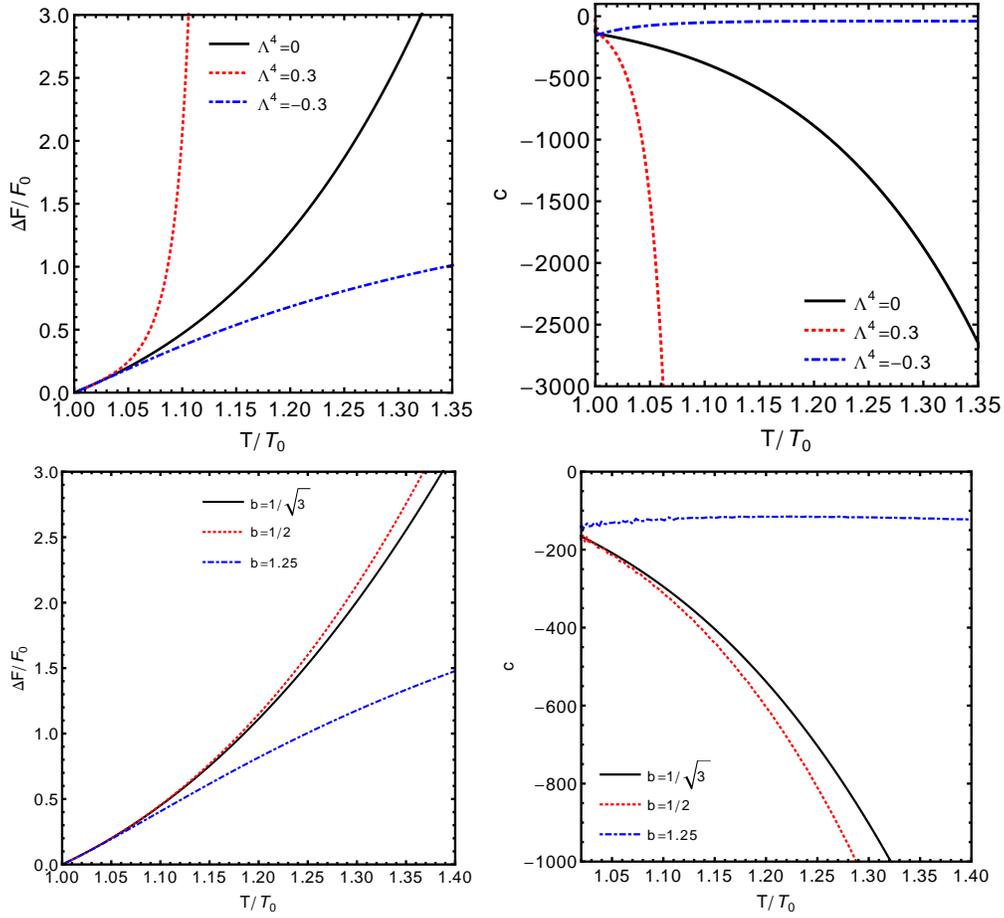

\begin{center}
\begin{tabular}{cc}
\epsfig{file=figure/DF_quartic.eps,height=6cm,angle=0,clip=true} & %
\epsfig{file=figure/specific_heat_quartic.eps,height=6cm,angle=0,clip=true}
\\ 
\epsfig{file=figure/cosh_DF.eps,height=6cm,angle=0,clip=true} & %
\epsfig{file=figure/cosh_specific_heat.eps,height=6cm,angle=0,clip=true}%
\end{tabular}%
\end{center}
\caption{Free energy (left panels) and specific heat (right panels) of the
dressed BBs as a function of the temperature. Top and bottom panels refer to
the theory with a quartic potential with $m_{s}^{2}=-2$ and with the
exponential potential~\eqref{pot}, respectively. }
\label{fig:energy}
\end{figure*}

\section{Charged solutions}

\label{sect:charged} In this section, we will extend the numerical results
previously obtained for neutral BBs in ES-AdS gravity to the case of finite
charge density, i.e to the case in which an EM field is present in the bulk.
We will focus our attention on models with exponential (\ref{pot}) or
quadratic (\ref{quadratic}) potential. We will discuss separately the cases
of: i) minimal gauge coupling $Z=1$; ii) exponential gauge coupling in the $%
U(1)-$symmetry preserving phase; iii) Minimal gauge coupling in the $U(1)-$%
symmetry breaking phase ($Z=1,\, Y\neq 0$).

\subsection{ Minimal gauge coupling}

In this section we will construct numerical BB solutions for the model (\ref%
{action}) with $Z=1,Y=0$ and the potential (\ref{pot}). As usual we discuss
separately extremal and finite temperature solutions.

\subsubsection{Extremal solutions}

Following the same approach as the one used for the case of electrically
neutral solutions, we look for numerical scalar-dressed BB's solutions
interpolating between an asymptotic AdS spacetime and a near-horizon
scale-covariant metric. Also in this case, the near-horizon behavior can be
captured by approximating the potential (\ref{pot}) in the $\phi \rightarrow
\infty $ region with the exponential form $V(\phi )=-(1/b^{2})e^{b\phi }$.
The field equations~\eqref{eqsEM1} -- \eqref{eqsEM3} give the scale
covariant solution, which in the dual QFT corresponds to hyperscaling
violation:%
\begin{eqnarray}
\lambda &=&\alpha _{0}\left( \frac{r}{r_{-}}\right) ^{w},\quad H=\left( 
\frac{r}{r_{-}}\right) ^{h},\quad \phi =\phi _{0}-\frac{b}{4}(w+2)\ln \left( 
\frac{r}{r_{-}}\right) ,\text{ \ \ }  \label{min} \\
w &=&2-4h=\frac{8-2b^{2}}{4+b^{2}},  \notag \\
\,\alpha _{0} &=&\frac{8e^{b\phi _{0}}r_{-}^{2}}{b^{2}w(w+2)},\quad \rho
^{2}=\frac{2e^{b\phi _{0}}(3w-2)}{b^{2}(w+2)},  \notag
\end{eqnarray}%
where $\rho $ is the charge density of the solution. The solution above,
together with the condition $\alpha _{0}>0$, restricts the parameter range
to $2/3<w<2$ (corresponding to $0<b^{2}<2$). We can exploit the symmetries
of the field equations to fix $L=1$ and $\phi _{0}=0$, leaving $r_{-}$ the
only free parameter. We immediately note an important feature of this
solution: in the limit $\rho \rightarrow 0$ it does not reduce to the
near-horizon solution (\ref{ex}) obtained in the electrically neutral case.
This means that the uncharged solution (\ref{ex}) and the electrically
charged solutions (\ref{min}) represent two disjoint classes of solutions.

\begin{figure*}[htb]
\begin{center}
\begin{tabular}{c}
\epsfig{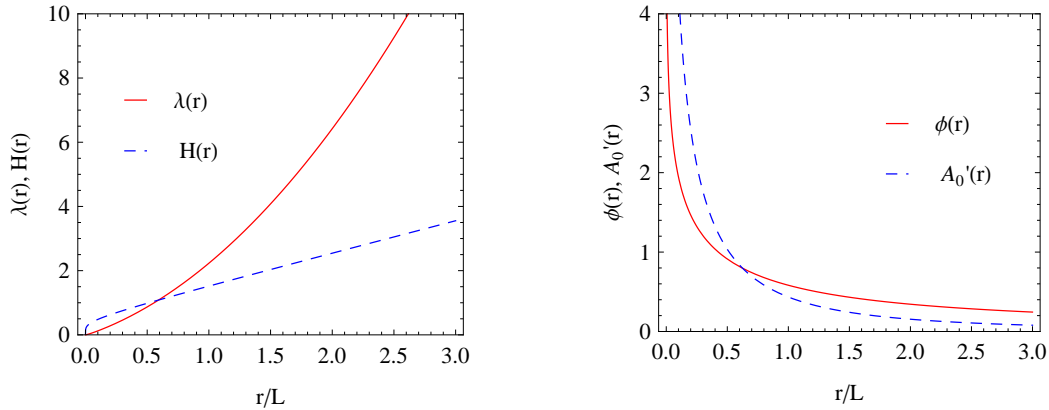}%
\end{tabular}%
\end{center}
\caption{Metric functions (left panel) and scalar and Maxwell field (right
panel), in the extremal case, for the potential (\protect\ref{pot}) with $b
=1$ and $r_{-}=1$.}
\label{minimal}
\end{figure*}

As usual, starting from this near-horizon scaling and imposing an AdS
behavior (\ref{bc}) at infinity, we have integrated numerically the field
equations for different values of the parameter $b$, finding numerical
solutions only for $b>1/2$. In Fig. \ref{minimal} we show the fields for $%
b=1 $. As expected, here we find in general $\lambda \neq H^{2}$, hence the
mass of the solution is nonvanishing and the degeneracy with the AdS vacuum
is removed.

\bigskip

\bigskip


\subsubsection{Finite-temperature solutions}

At variance with the extremal case, the charge of finite temperature
solutions is a free parameter and the uncharged case is obtained setting $%
\rho=0$. Using a straightforward extension of the numerical integration
previously discussed, we can construct finite-temperature solutions at fixed
charge density $\rho$. Some examples are shown in Fig.~\ref%
{fig:cosh_profiles_rho} for the potential~\eqref{pot} with $b=1$. In the
left panel we show the radial profiles of the fields, in the central panel
we show the function $O_2=O_2(O_1)$ for different values of $\rho$, and in
the right panel we shown the difference between the free energy of the
dressed solution and that of a RN BB with same radius and same charge.
Notice that, as already stressed, in the charged case the boundary
conditions can be arbitrarily chosen. In particular, one can also choose
conformal boundary conditions of the form $O_{1}=0$. However, in the case at
hand we have found that such conditions do not allow for scalar-dressed BBs.

Similarly to the uncharged case, these dressed solutions are always
energetically disfavored with respect to the undressed ones.

\begin{figure*}[htb]
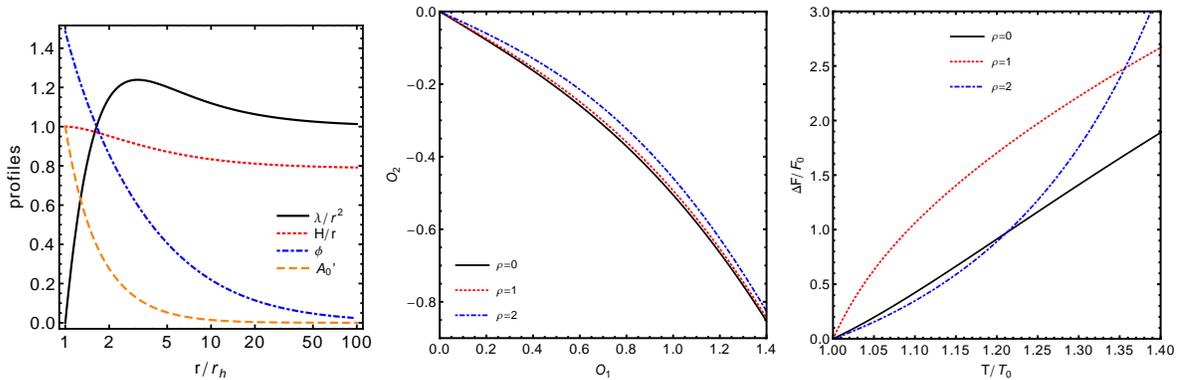

\begin{center}
\begin{tabular}{ccc}
\epsfig{file=figure/cosh_profiles_rho2.eps,height=5cm,angle=0,clip=true} & %
\epsfig{file=figure/cosh_O2O1_rho.eps,height=5cm,angle=0,clip=true} & %
\epsfig{file=figure/cosh_DF_rho.eps,height=5cm,angle=0,clip=true}%
\end{tabular}%
\end{center}
\caption{Left panel: profiles for the metric coefficients, scalar field and
EM potential as functions of the (nonrescaled) coordinate $r$ for the
potential~\eqref{pot} with $b=1$ and different values of $\protect\rho$.
Central panel: the functions $O_2=O_2(O_1)$ for different values of $\protect%
\rho$ and $b=1$. Right panel: difference between the free energy of the
dressed solution and that of a RN BB with same radius and same charge.}
\label{fig:cosh_profiles_rho}
\end{figure*}

\bigskip

\subsection{Nonminimal gauge coupling}

As an example of a model with nonminimal gauge coupling we consider here the
model discussed in Ref. \cite{Cadoni:2009xm}. The gauge coupling $Z$ and the
potential $V $ are \cite{Cadoni:2009xm}

\begin{equation}
Z(\phi )=2Z_{0}\cosh a\phi ,\text{ \ \ \ \ \ }\,V(\phi )=-2V_{0}\cosh b\phi .
\label{nonmin}
\end{equation}%
In the IR ($\phi \rightarrow \infty $) both the gauge coupling $Z$ and the
potential $V$ behave exponentially. The model therefore belongs to the wide
class of EHTs that flow to a hyperscaling violating phase in the IR \cite%
{Charmousis:2009xr,Cadoni:2009xm,Goldstein:2009cv,Dong:2012se,Gouteraux:2012yr}%
. The extremal solution of the field equations in the near-horizon
approximation is the scale-covariant metric \cite{Cadoni:2009xm}: 
\begin{eqnarray}
\lambda &=&\alpha _{0}\left( \frac{r}{r_{-}}\right) ^{w},\quad H=\left( 
\frac{r}{r_{-}}\right) ^{h},\quad \phi =\phi _{0}-\xi \ln \left( \frac{r}{%
r_{-}}\right) ,\text{ \ \ }  \label{p1} \\
\xi &=&\frac{4(a+b)}{4+(a+b)^{2}},\text{ \ \ }w=2-4ch,\text{ \ \ \ \ }c=%
\frac{b}{a+b},  \notag \\
\,\alpha _{0} &=&\frac{2V_{0}e^{b\phi _{0}}r_{-}^{2}}{(w+2h)(w+2h-1)},\quad 
\frac{\rho ^{2}}{Z_{0}}e^{-a\phi _{0}}=\frac{2V_{0}e^{b\phi _{0}}(2-2h-b\xi )}{%
(w+2h)}.  \notag
\end{eqnarray}

In Ref. \cite{Cadoni:2009xm} $T=0$ global solutions interpolating between
the near-horizon hyperscaling violating metric (\ref{p1}) and the asymptotic
AdS$_{4}$ geometry have been constructed numerically. Furthermore, numerical
finite-temperature solutions have been found and their properties have been
discussed in detail. In particular, it has been shown that below a critical
temperature $T_{c}$ the system undergoes a phase transition. The scalar
dressed BB solution becomes energetically preferred with respect to the RN
BB.

The results of Ref. \cite{Cadoni:2009xm} fully confirm the general results
of Sect. \ref{sect:extremal}. The finite charge density removes the
degeneracy of the $T=0$ solution we have in the uncharged case. Comparing
the charged solution (\ref{p1}) with the neutral solution (\ref{ex}) one
easily realizes that although the IR behavior of the two solutions belongs
to the same class (hyperscaling violating) the critical exponents change.
Moreover, the nonminimal coupling between the scalar field and the Maxwell
field is such that the energy of the extremal scalar-dressed solution is
smaller than that of the RN-solution. This determines an IR quantum phase
transition between the AdS$_{2}\times R^{2}$ near-horizon geometry of the RN
BB and the near-horizon scale covariant geometry (\ref{p1}). In the dual QFT
this corresponds to a phase transition between a conformal and a
hyperscaling violating fractionalized phase.

Because the thermodynamical properties of the system at small temperature
are essentially determined by the $T=0$ quantum phase transition, this also
explains why the hyperscaling violating phase is stable at small
temperature, below $T_{c}$.

The near-horizon solutions for the charged BB (\ref{min}) and (\ref{p1})
depend on the same IR length-scale $r_{-}$ as the neutral BB solution (\ref%
{ex}). However, in the charged case the scaling transformations under which
the metric part of the solutions is scale-covariant, changes not only 
the constant mode of the scalar 
$\phi_{0}$ but also the charge density $\rho$. Thus, changing the IR scale $r_{-}$
corresponds to a flow of the charge density $\rho$. As noticed already in
Ref. \cite{Gouteraux:2012yr}, this is an irrelevant deformation along the
hyperscaling violating critical line.

It is also interesting to notice the different role played in the quantum
phase transition by the finite charge density and the nonminimal gauge
coupling. The finite charge density lifts the degeneracy of the $T=0$ vacuum
and changes the values of the critical exponents of the hyperscaling
violating solution, but it is by itself not enough to make the hyperscaling
violating phase energetically competitive with respect to the conformal AdS$%
_{2}\times R^{2}$ phase. Indeed, in the case of minimal gauge couplings
discussed in the previous subsection, the energy of extremal RN-BBs is
lesser than the energy of the scalar-dressed $T=0$ solution. It is the non
minimal coupling between the gauge and the scalar field that makes the
extremal RN solution energetically disfavored with respect to the extremal
scalar-dressed solution.

\subsubsection{Hyperscaling violation and critical exponents}

In the case of a potential behaving exponentially in the IR the
near-horizon, extremal solutions are scale-covariant for both zero or
finite charge density and for both minimal or nonminimal gauge couplings. On
the other hand, the critical exponents are affected by switching on a finite
charge density. In particular in the case of charged solutions we will
always have $z\neq 1$.

In the minimal case, after a redefinition of the radial coordinate and a
rescaling of the coordinates, the metric (\ref{min}) reads: 
\begin{equation*}
ds^{2}=r^{2}\left( -r^{\frac{2(3w-2)}{2-w}}dt^{2}+dr^{2}+dx_{i}^{2}\right) ,
\end{equation*}%
from which we can easily extract the critical parameters: 
\begin{equation*}
\theta =4,\text{ \ }z=\frac{2(2-2w)}{2-w}.
\end{equation*}%
We note immediately that the hyperscaling violation exponent $\theta $ is a
(positive) constant, independent from the parameters of the potential. The
range of $w$ implies $z<1$, which is in agreement with the NEC conditions.
Indeed the latter impose, for this values of $\theta $ and $z$, the
conditions $z>2$ or $z<1$. Moreover we note that for $1<w<2,$ $z$ is
negative.

On the other hand, for $2/3<w<1$ (corresponding to $0<z<1$), the free energy
scales with a negative exponent: 
\begin{equation*}
F\sim T^{\frac{2-\theta +z}{z}}=T^{\frac{w}{2w-2}},
\end{equation*}%
which implies an instability of the corresponding phase and a negative
specific heat.

Finally, we consider the case of a nonminimal gauge coupling given by Eq. (%
\ref{nonmin}). The critical exponents can be read off from Eq. (\ref{p1}),
after a reparametrization of the radial coordinate which puts the metric in
the form (\ref{dwh}). We have 
\begin{equation*}
\theta =\frac{4c}{2c-1}\ \ ,\ \ z=\frac{2c(2-2w)}{(2c-1)(2-w)},
\end{equation*}%
while the free energy scales as: 
\begin{equation}
F\sim T^{\frac{2-\theta +z}{z}}=T^{\frac{(2c-1)w+2-2c}{c(2w-2)}}.
\end{equation}

\subsection{Symmetry-breaking phase}

\label{sect:sbreak} As an example of a model having a $U(1)$%
-symmetry-breaking phase we consider here the model discussed in Ref. \cite%
{Hartnoll:2008vx,Hartnoll:2008kx, Horowitz:2008bn,Horowitz:2009ij}, which
gives the simplest realization of holographic superconductors. The gauge
coupling is minimal, while the potential $V$ and the function $Y(\phi )$ in
the action (\ref{action}) are quadratic \cite{Horowitz:2009ij}: 
\begin{equation*}
Z(\phi )=1,\,\text{\ \ \ \ \ }V(\phi )=\frac{m_{s}^{2}}{2}\phi ^{2},\,\ \ \
\ \ Y(\phi )=q^{2}\phi ^{2},
\end{equation*}%
where $q$ is the electric charge of the complex scalar field whose modulus
is $\phi $.

The metric and scalar field associated to the $T=0$ solution of the field
equations in the near-horizon approximation are given as in the neutral case
discussed in Sect. \ref{sect:quadratic}, i.e. by Eq. (\ref{hr}), whereas the
EM potential is $A_{0}=\phi _{0}\hat{r}^{\beta }(-\log \hat{r})^{1/2}$ with $%
2\beta =-1\pm (1-48q^{2}/m_{s}^{2})^{1/2}$. Numerical, extremal solutions
interpolating between the near-horizon solution (\ref{hr}) and AdS$_{4}$
have been constructed for $q^{2}>|m^{2}_{s}|/6$ in Ref. \cite%
{Horowitz:2009ij}. Numerical finite-temperature solutions have been also
considered~\cite{Hartnoll:2008vx,Hartnoll:2008kx, Horowitz:2008bn}. In
particular, it is well known that below a critical temperature the
superconducting phase (corresponding in the bulk to the scalar-dressed BB
solution) becomes energetically preferred.

The results of Ref. \cite{Hartnoll:2008vx,Hartnoll:2008kx,
Horowitz:2008bn,Horowitz:2009ij} for the holographic superconductors fully
confirm our general results of Sect. \ref{sect:extremal}. The finite charge
density removes the degeneracy of the $T=0$ solution in the uncharged case.
Moreover, the nonvanishing coupling function $Y$ gives a mass to the $U(1)$
gauge field and makes the extremal scalar-dressed solution energetically
competitive with respect to the RN-solution. The system represents an IR
quantum phase transition between the AdS$_{2}\times R^{2}$ near-horizon
geometry of the RN BB and the near-horizon geometry (\ref{hr}). In the dual
QFT this corresponds to the superconducting phase transition \cite%
{Hartnoll:2008vx,Hartnoll:2008kx, Horowitz:2008bn,Horowitz:2009ij}, which
occurs below the critical temperature.

Similarly to the nonminimal case, also here the finite charge density and
the nonvanishing function $Y$ play a very different role. The finite charge
density simply lifts the degeneracy of the $T=0$ vacuum we have in the
uncharged case. But it is the coupling between the scalar field and the EM
potential $A_{0}$ that causes the superconducting phase transition to occur
at the critical temperature. It is also interesting to notice that in this
case the finite charge density does not change the metric (and scalar) part
of the IR solution, which is determined by the near-horizon solution and it
is described as in the EM neutral case by Eq.~(\ref{hr}).


\section{Concluding remarks}

\label{sect:concluding} In this paper we have discovered several interesting
features of scalar condensates in EHTs, which may be relevant for
understanding holographic quantum phase transitions. In particular, we have
shown that for zero charge density the ground state for scalar-dressed,
asymptotically AdS, BBs must be degenerate with the AdS vacuum, must be
isolated from the finite-temperature branch of the spectrum and must satisfy
conformal boundary conditions for the scalar field. This degeneracy is the
consequence of a cancellation between a gravitational positive contribution
to the energy and a negative contribution due to the scalar condensate. When
the scalar BB is sourced by a pure scalar field with a potential behaving
exponentially in the IR, a scale is generated in the IR. 

Switching on a finite charge density $\rho$ for the scalar BB, the
degeneracy of the ground state is removed, the ground state is not anymore
isolated from the continuous part of the spectrum and the  flow of 
the IR scale typical
of hyperscaling violating geometries determines a flow of   $\rho$. Depending on
the gauge coupling between the bulk scalar and EM fields, the new ground
state may be or may not be energetically preferred with respect to the extremal
RN-AdS BB. We have also explicitly checked these features in the case of
several charged and uncharged scalar BB solutions in theories with minimal,
nonminimal and covariant gauge couplings. In the following subsections we
will briefly discuss the consequences our results have for the dual QFT and
for quantum phase transitions.

\subsection{Dual QFT}

One striking feature of the uncharged scalar BB solutions we discussed is
that the boundary conditions for the scalar field are either determined by
the symmetries (for the ground state) or by the dynamics (for
finite-temperature solutions). Because the only free function entering the
model is the scalar potential $V(\phi)$, this means that the information
about boundary conditions for the scalar field is entirely encoded in the
symmetries of the field equations and in $V$. Since the scalar field drives
the holographic renormalization group flow, this fact has some interesting
consequence for the dual QFT.

We have seen in Sect.~\ref{sect:extremal} that in the case of zero charge
density the ground state for the scalar BB must be characterized by
conformal boundary conditions. From the point of view of the dual QFT this
corresponds to a multi-trace deformation of the Lagrangian of the CFT. This
is a relevant deformation, associated to a relevant operator, which will
produce a renormalization-group flow from an UV CFT to an IR QFT. The nature
of the IR QFT is entirely determined by the self-interaction potential, $V(\phi )$. 
In the case of the quartic potential (\ref{quartic})
--~which is characterized by two extrema~-- the IR QFT has the form of a
further CFT. In the case of the exponential potential (\ref{moda}) the IR
QFT is characterized by hyperscaling violation. In the case of the quadratic
potential (\ref{quadratic}) the characterization of the IR QFT is much less
clear because of the absence of scaling symmetries.

The characterization of the dual QFT at finite temperature is much more
involved. In this case we have generically nonconformal boundary conditions
for the scalar field and the asymptotic AdS isometries are broken.
Nonetheless, an asymptotic time-like killing vector exists and both the UV
and the IR QFT should admit a description in terms of multi-trace
deformations of a CFT.

On the other hand we have shown that the ground state and finite$-T$ states
are not continuously connected. This means that we are dealing here with two
different disjoint sets of theories.

This picture changes completely when one adds a finite charge density. Now the boundary
conditions for the scalar field can be arbitrarily chosen, for instance in
the form of the usual conformal Neumann or Dirichlet boundary conditions.
Thus, in the case of finite charge, we have the usual description borrowed
from the AdS/CFT correspondence with single trace operators dual to the
scalar field.

\subsection{Scalar condensate and quantum criticality}

The results of this paper improve our understanding of quantum critical
points in EHTs. In particular they shed light on the phase structure of
these critical points proposed in Ref. \cite{Gouteraux:2012yr} and on their
stability.

The degeneracy of the ground state for uncharged BBs simply means that at
zero charge density the hyperscaling violating critical point (or line) and
the hyperscaling-preserving critical point have the same energy. The
potential $V$ for the scalar field determines completely the scaling
symmetry and the critical exponents of the hyperscaling violating critical
point. The renormalization group flow from the UV conformal fixed point into
the IR introduces an emergent IR scale. Changing this IR scale 
produces a flow of the constant mode of the scalar field. As already noted 
in Ref.~\cite{Gouteraux:2012yr}, the
presence of this arbitrary scale implies that hyperscaling violating
critical points appear as critical lines rather than critical points. On the
other hand, for scalar-dressed BBs the ground state is isolated from the
finite$-T$ part of the spectrum and the states at finite temperature are
always energetically disfavored with respect to the SAdS BB. Thus, at zero
charge density there is no phase transition between the hyperscaling
preserving phase and the hyperscaling violating phase.

Considering charged scalar BBs, i.e. introducing a finite charge density $%
\rho $ in the dual QFT, generates several effects. First of all the
degeneracy of the ground state is lifted and the ground state is not anymore
isolated from the $T>0$ continuous branch of the spectrum. The change 
of the IR scale
typical of hyperscaling violating critical lines now also produces a flow 
of the charge density $\rho $.
Although the critical exponents are modified by the presence of a finite
charge density (for instance the dynamical critical exponent $z$ becomes $%
\neq 1$) the scaling symmetries characterizing the critical point are very
similar to those we have in the case of $\rho =0$. The similarity between
the ground state geometries in the $\rho \neq 0$ and $\rho =0$ case is even
more striking in the case of covariant gauge coupling (the case of
holographic superconductors). In this latter case the metric and scalar part
of the near-horizon solution is exactly the same for $\rho \neq 0$ and $\rho
=0$.

The stability of the hyperscaling violating critical line is a far more
involved question. It turns out that it depends crucially on the coupling
between the scalar condensate and the EM field, i.e. on the two coupling
functions $Z(\phi)$ and $Y(\phi)$ in the action (\ref{action}). In all cases
that we have considered with a minimal gauge coupling $Z=1$, and in absence of $U(1)$ symmetry
breaking ($Y=0$), the hyperscaling preserving phase is
always energetically preferred with respect to the hyperscaling violating
one. In this case, an IR phase transition between
hyperscaling-preserving phase and the hyperscaling violating phase does not occur.

Conversely, in the two cases of a nonminimal gauge coupling behaving
exponentially in the IR ($Z\sim e^{a\phi}$, $Y=0$) and covariant gauge
coupling ($Z=1, \, Y\sim \phi^{2}$), the hyperscaling violating phase is
energetically preferred. This gives, respectively, the IR phase transitions
between the hyperscaling preserving phase and the hyperscaling violating
phase found in Ref.~\cite{Cadoni:2009xm} and the well-known superconducting
phase transition of Ref. \cite{Hartnoll:2008vx,Hartnoll:2008kx,
Horowitz:2008bn,Horowitz:2009ij}. On the other hand, considering charged BBs at finite
temperature, the critical temperature of the phase transition between the
hyperscaling-preserving/hyperscaling violating phases is settled by the
charge density $\rho$ \cite{Cadoni:2009xm}, i.e by the IR emergent scale
typical of the hyperscaling violating critical line.

Summarizing, our results strongly indicate that for EHTs described by (\ref%
{action}) the three coupling functions $V(\phi ),\,Z(\phi ),\,Y(\phi )$
determine different features of holographic quantum critical points. The
self-interaction potential $V(\phi )$ determines the scaling symmetries but
not the stability of hyperscaling violating phases. Conversely $Z$ and $Y$
are crucial in determining the stability, the breaking of the $U(1)$
symmetry and the characterization as fractionalized or cohesive of the
hyperscaling violating phase.



\begin{acknowledgements}
P.P. acknowledges financial support provided by the European
Community 
through the Intra-European Marie Curie contract aStronGR-2011-298297
and by FCT-Portugal through projects
PTDC/FIS/098025/2008, PTDC/FIS/098032/2008 and CERN/FP/123593/2011.

M.S. gratefully acknowledges Sardinia Regional Government for the
financial support of his PhD scholarship (P.O.R. Sardegna F.S.E.
Operational Programme of the Autonomous Region of Sardinia, European
Social Fund 2007-2013 - Axis IV Human Resources, Objective I.3, Line
of Activity I.3.1).
\end{acknowledgements}

\appendix

\section{Uncharged perturbative solutions in the small scalar-field limit}

\label{sect:appendix} 
In the neutral case, it is possible to construct analytical BB solutions in
the small scalar-field limit perturbatively, i.e. expanding the solution as
follows 
\begin{eqnarray}
\lambda(r)&=&\frac{r^2}{L^2}-\frac{M}{2r}+\epsilon^2\lambda_2(r)\,, \\
H(r)&=&r+\epsilon^2H_2(r)\,, \\
\phi(r)&=&\epsilon\phi_1(r)\,.
\end{eqnarray}
where $\epsilon$ is a book-keeping parameter of the expansion. The solution
for the scalar field can be obtained by solving the scalar equation at first
order. The regular solution, can then be inserted into the Einstein
equations that, to second order, can be solved for $\lambda_2$ and $H_2$.

Let us start with the $T=0$ AdS$_4$ vacuum, i.e. we set $M=0$ in the
equations above. To second order in the scalar field, the solution reads 
\begin{eqnarray}
\lambda(r)&=&\frac{r^2}{L^2}+\left(-\frac{O_1^2}{4 L^2}-\frac{O_2^2}{6 L^2
r^2}+\frac{2 r C_1}{L^2}+\frac{C_2}{r}\right) \epsilon^2+\mathcal{O}%
(\epsilon^4)\,, \\
H(r)&=&r+\left(-\frac{O_2^2}{12 r^3}-\frac{O_1 O_2}{6 r^2}-\frac{O_1^2}{8 r}%
+C_1+r C_2\right) \epsilon^2+\mathcal{O}(\epsilon^4)\,, \\
\phi(r)&=&\epsilon\left(\frac{O_1 }{r}+\frac{O_2}{r^2}\right)+\mathcal{O}%
(\epsilon^3)\,,
\end{eqnarray}
where $C_i$ are integration constants. This is a solution for the classes
of potentials presented in the main text. Although not presented, the
solutions can be obtained in closed form at least to fourth order. The
constant $C_1$ can be set to zero by performing a coordinate translation
such that the asymptotic form of the metric reads as in Eq.~%
\eqref{expansion_inf} with $C_2=-m_0/2$ being related to the metric
contribution to the gravitational mass and after a rescaling $H\to
H/(1+\epsilon^2 C_2)$, which can be performed by rescaling the the
transverse coordinates. Interestingly, there exists an event horizon, so the
solution represents a BB endowed with a scalar field. Let us consider two
cases separately: $O_2=0$ and $O_2=O_2(O_1)$ (without loss of generality, we
assume $O_2\geq0$). For the latter case, the horizon is located at 
\begin{equation}
r_h=\frac{\sqrt{O_2}\epsilon}{6^{1/4}}+\frac{\sqrt{3} m_0 }{4\sqrt{2} O_2}%
\epsilon+\frac{3^{1/4} \left(4 O_1^2 O_2^2-3 m_0^2\right)}{ 2^{3/4}32
O_2^{5/2}}\epsilon^{3/2}+\mathcal{O}(\epsilon^{5/2})\,,
\end{equation}
and, to first order, the temperature of the solution is 
\begin{equation}
T=\frac{\sqrt{O_2} \sqrt{\epsilon}}{6^{1/4} \pi }\,.
\end{equation}
On the other hand, if $O_2=0$, the horizon and the temperature read 
\begin{eqnarray}
r_h&=&\frac{m_0^{1/3} \epsilon^{2/3}}{2^{1/3}}+\frac{O_1^2 \epsilon^{4/3}}{
2^{2/3} 6 m_0^{1/3}}\,, \\
T&=&\frac{3 m_0^{1/3} \epsilon^{2/3}}{2^{1/3} 4\pi }+\frac{O_1^4 \epsilon^2}{%
96 m_0 \pi }
\end{eqnarray}

In general, these solutions describe a BB whose horizon shrinks to zero in
the $\mathcal{O}_i\to0$ limit. The total mass of the BB is given by Eq.~%
\eqref{mass} and it coincides with $m_0$ when $O_2=0$. It is interesting to
compare the free energy of this solution with that of a SAdS BB at the same
temperature. When $O_2\neq0$, we obtain 
\begin{equation}
F-F_0=\frac{37 (\epsilon O_2)^{3/2}}{6^{3/4} 27}+\mathcal{O}(\epsilon^2)\,,
\end{equation}
so that $F>F_0$ for any $O_2\neq0$ and the dressed solution is always
energetically disfavored. Note that this result is valid for any boundary
condition $O_2=O_2(O_1)\neq0$ and for any scalar potential whose expansion
reads $V\sim-6/L^2-\phi^2/L^2$. On the other hand, if $O_2=0$, $F=F_0$ to
second order in $O_1$, so that the two solutions are degenerate.

Finally, we can adopt the same technique to construct perturbative solutions
of the SAdS BB at finite temperature. At first order, the general solution
of the scalar field equation reads 
\begin{equation}
\phi_1=\alpha P_{-1/3}\left[{r^3}/(L^2M)-1\right]+\beta Q_{-1/3}\left[{r^3}%
/(L^2M)-1\right]\,,
\end{equation}
where $P_n$ and $Q_n$ are Legendre functions of order $n$ and $\alpha$ and $%
\beta$ are integration constants. Imposing regularity at the horizon $%
r_h=(2L^2M)^{1/3}$ requires $\beta=0$. In principle, this solution can be
inserted in the Einstein equation in order to obtain two equations for $%
H_2(r)$ and $\lambda_2(r)$. Unfortunately, these equations do not appear to
be solved in closed form.


\bigskip

\bigskip

\bigskip

\bigskip

\bigskip

\bigskip

\bigskip

\bigskip

\bibliography{infrared}

\end{document}